\documentclass{article}

\usepackage{iftex}
\ifPDFTeX
  \errmessage{This document requires XeLaTeX or LuaLaTeX because it uses fontspec/xeCJK}
\fi

\usepackage{fontspec}
\usepackage{xeCJK}

\setCJKsansfont{FandolHei-Regular.otf}[
  BoldFont = FandolHei-Bold.otf
]
\setCJKmonofont{FandolFang-Regular.otf}

\usepackage[top=30mm,left=30mm,right=30mm,bottom=30mm,heightrounded]{geometry}
\usepackage{parskip}

\usepackage[dvipsnames,svgnames,x11names]{xcolor}
\definecolor{lightgray}{gray}{0.97}

\usepackage{graphicx}
\usepackage{adjustbox}
\usepackage{wrapfig}
\usepackage{pdflscape}
\usepackage{float}
\usepackage{caption}
\usepackage{subcaption}

\usepackage{booktabs}
\usepackage{array}
\usepackage{tabularx}
\usepackage{tabulary}
\usepackage{longtable}
\usepackage{multirow}
\usepackage{makecell}
\usepackage{threeparttable}
\usepackage{threeparttablex}
\usepackage{colortbl}
\usepackage{ragged2e}

\newcolumntype{L}[1]{>{\raggedright\arraybackslash}p{#1}}

\usepackage{listings}
\usepackage[most]{tcolorbox}

\lstdefinestyle{promptstyle}{
  basicstyle=\ttfamily\small,
  backgroundcolor=\color{lightgray},
  breaklines=true,
  columns=fullflexible,
  keepspaces=true,
  showstringspaces=false
}

\newtcblisting{promptbox}{
  listing only,
  colback=lightgray,
  colframe=black!20,
  boxrule=0.3pt,
  arc=2pt,
  left=6pt,
  right=6pt,
  top=6pt,
  bottom=6pt,
  listing options={style=promptstyle}
}

\usepackage{etoolbox}
\usepackage{xpatch}

\makeatletter
\patchcmd\longtable{\par}{\if@noskipsec\mbox{}\fi\par}{}{}
\makeatother

\IfFileExists{footnotehyper.sty}
  {\usepackage{footnotehyper}}
  {\usepackage{footnote}}

\makesavenoteenv{longtable}

\usepackage[style=apa,backend=biber]{biblatex}
\usepackage[page,toc,title]{appendix}
\usepackage{minitoc}

\usepackage{authblk}
\usepackage{orcidlink}

\usepackage{xurl}
\usepackage{hyperref}
\usepackage{bookmark}

\hypersetup{
  pdftitle={Beyond Killer Robots},
  colorlinks=true,
  linkcolor=black,
  citecolor=black,
  urlcolor=Blue,
  filecolor=Maroon,
  pdfcreator={LaTeX}
}

\providecommand{\tightlist}{%
  \setlength{\itemsep}{0pt}
  \setlength{\parskip}{0pt}
}

\title{Beyond Killer Robots: General AI Attitudes and Public Support for Military AI in Nine Countries}

\author[1,2]{Andreas Jungherr \orcidlink{0000-0003-2598-2453}}
\author[2]{Antonia Schlude}
\author[3]{Adrian Rauchfleisch \orcidlink{0000-0003-1232-083X}}

\affil[1]{University of Bamberg}
\affil[2]{Bavarian Research Institute for Digital Transformation (bidt)}
\affil[3]{National Taiwan University}

\date{\today}

\begin{document}

\doparttoc
\faketableofcontents

\maketitle


\begin{abstract}

AI-enabled military systems are a fixture of modern military conflict.
Applications vary from autonomous drones for surveillance and attack to
AI-supported target selection. The importance of AI for modern conflict
shows also in public disputes between governments and technology
companies over the conditions for military access to frontier AI. Both
military uses and government attempts at enabling and steering them
happen before a backdrop of public opinion, yet we still know little
about how people think about military AI. Drawing on a preregistered
survey of 9,000 respondents in nine countries, including China, Germany,
and the United States, we examine whether support for military AI is
shaped primarily by general attitudes toward AI, principled opposition
to lethal autonomy, or foreign-policy and geopolitical orientations.
Across six military AI scenarios that vary in lethality and human
control, respondents who view AI as beneficial are substantially more
supportive of military AI. Hawkish respondents are also more supportive.
By contrast, principled opposition to lethal autonomy is not broadly
associated with the full index, but is related to the application of
fully autonomous lethal force. Contrary to our expectation, perceived AI
risks are positively associated with support. Cross-national differences
are moderate and broadly consistent with geopolitical context. Overall,
public opinion toward military AI appears conditionally permissive.
Publics are not categorically opposed to various military uses of AI.
Instead, unease is concentrated around fully autonomous lethal force.
\end{abstract}



\begin{refsection}

\section{Introduction}\label{introduction}

By 2026, military AI has become a fixture of modern military conflict.
In Ukraine's defense against Russia's war of aggression, autonomous
drones and ground robots have become an important feature of the
conflict and a source of tactical advantage
\autocite{Bondar:2025aa,Melkozerova:2026ab,Rees:2025aa}. The US air
strikes against Iran in 2026 reportedly were based on AI-enabled target
selection, with reports attributing these uses the ability for the US to
hit an unprecedented number of targets in a short amount of time.
Reports of civilian casualties, however, reinforced concerns about
algorithmic decision support in combat
\autocite{Airwars-Staff:2026aa,Copp:2026aa,The-Editorial-Board:2026aa}.
At the same time, the Pentagon entered a public dispute with the
frontier AI lab \emph{Anthropic} over restrictions the company sought to
place on military uses of its systems
\autocite{Frenkel:2026aa,Horowitz:2026aa,Thompson:2026aa}, surfacing a
tension between civilian AI governance and state demands to sovereignly
pursue security policy \autocite{Wong:2026aa}. Military AI is no longer
only a speculative concern for arms-control specialists or a
technological frontier for defense planners and tech-industrialists. It
has become a visible object of international security politics and with
this, turned into an object for public opinion.

Public opinion provides the backdrop before which military
organizations, technology companies, and civil society act.\footnote{For
  public opinion as an environment for the conduct of war and security
  policy, see \textcite{Baum:2015aa}; \textcite{Berinsky:2009aa};
  \textcite{Page:1992aa}; \textcite{Holsti:1996aa}.} It can enable
military AI through a ``permissive consensus''
\autocite{Holsti:1996aa,Hooghe:2009aa} or restrict it through targeted
norm entrepreneurship by civil-society networks
\autocite{Bode:2022aa,Carpenter:2014ab,Keck:1998aa,Rosert:2021aa}. So
far, however, research providing a systematic overview of how people
think about military AI and what other considerations relate to this
view is limited. Most public and academic debate about military AI has
focused on the use case of lethal autonomous weapons (LAWs), often
evocatively framed as ``killer robots'', and the associated moral
problem of machines making life-and-death decisions. This is an
important concern. But fully autonomous lethal systems are only an edge
case among a much broader set of military uses of AI.

We use ``military AI'' to refer to AI-enabled systems and capabilities
developed for or deployed by the military. The term covers a broad set
of uses \autocite{Horowitz:2018aa,Scharre:2018aa}. This includes
intelligence and surveillance, logistics and back-office operations,
command and decision support, cyber and electronic warfare, targeting,
and autonomous lethal systems. These applications differ in two
politically and ethically important respects: whether they are directly
involved in lethal force and how much decision-making authority they
delegate from humans to machines. These dimensions provide a useful way
to structure our analysis of how people react to different uses of
military AI. Concerns about lethal autonomous weapons might only apply
to a small subset of military AI, or, conversely, might come to dominate
considerations of military AI more generally.

We examine public opinion on military AI with original survey data from
9,000 respondents in China, Finland, France, Germany, Italy, Spain,
Taiwan, the United Kingdom, and the United States, collected for us by
\emph{Ipsos} in January 2026. The study was preregistered before data
collection.\footnote{Preregistration:
  https://osf.io/m8r5n/overview?view\_only=64d58c1e2b254a27bad6cc418a2e1105}
Our dependent variable is a six-item battery that crosses lethality and
degree of automation: non-lethal decision support, non-lethal
human-in-the-loop use, non-lethal autonomous use, lethal decision
support, lethal human-in-the-loop use, and fully autonomous lethal
force. This design allows us to compare support for military AI across
applications that vary in ethical salience, operational function, and
the degree to which meaningful human control is at stake.

The article proposes three possible explanations of public opinion on
military AI: The first expects people to treat military AI predominantly
as an extension of AI more generally. This would make military AI an
extension of technologies people already encounter through chatbots,
recommender systems, image generators, public-sector automation, and
debates over civilian AI governance. This view builds on public-opinion
research on AI as a general-purpose technology. It expects attitudes
toward military AI to follow general beliefs about AI's societal
benefits and risks.\footnote{For public attitudes toward AI as a
  general-purpose technology, see \textcite{Zhang:2019ab};
  \textcite{Bao:2022aa}; \textcite{Kreps:2023aa};
  \textcite{Jungherr:2024aa}.}

The second explanation expects people to see different military uses of
AI through the frame of lethal autonomy and on moral problem of
delegating life-and-death decisions to machines. Here we build on
security ethics and autonomous-weapons scholarship.\footnote{For the
  ethics of lethal autonomous weapons and meaningful human control, see
  \textcite{Sparrow:2007aa}; \textcite{Sharkey:2012aa};
  \textcite{Asaro:2012aa}; \textcite{Crootof:2016aa}.} This explanation
expects concerns about human control, responsibility, and accountability
to travel from the specific use case of lethal autonomous weapons to
military AI more broadly.

The third explanation expects people to approach military AI less as a
question about AI than a question about military power and the role of
force in international disputes. Here we build on public-opinion
research in foreign policy and international security.\footnote{For
  foreign-policy public opinion and the structure of hawk--dove
  orientations, see \textcite{Wittkopf:1990aa};
  \textcite{Hurwitz:1987aa}; \textcite{Kertzer:2014aa};
  \textcite{Mader:2025aa}.} In this view, support for military AI should
depend on hawk-dove orientations, views of the international order, and
trust in major powers, considerations featuring prominently in earlier
empirical work on public attitudes toward autonomous weapons and
military AI.\footnote{For empirical research on public attitudes toward
  autonomous weapons and military AI, see \textcite{Horowitz:2016aa};
  \textcite{Zwald:2025aa}; \textcite{Horowitz:2025aa};
  \textcite{Lillemae:2025aa}.}

Each explanation suggests a different source for public support or
concern and thereby foundations of democratic constraint. General AI
beliefs would make civilian AI experiences and broader AI discourse
politically relevant for military AI. The autonomous-weapons frame would
make principled opposition to lethal autonomy and concern for meaningful
human control central. The military-power frame would instead point to
hawkishness, perceptions of international order, and the salience of
particular security relationships. The empirical question our article
answers is how much weight each of these explanations carries when they
are examined together and across different national contexts.

Our findings show that people seem not to differentiate between most
uses of military AI. The main exception is automated lethal force.
Support is consistently higher for non-lethal than lethal uses, and it
falls as automation increases. Fully autonomous lethal force is the
least-supported application in every country. This points to the
continuing importance public opinion puts on human control over lethal
force. It does not mean, however, that attitudes toward military AI
generally follow a principled opposition to lethal autonomy. In the
broader six-item battery, perceived AI benefits are the strongest
predictor of support. Hawk-dove orientations follow as a second
independent predictor. Principled opposition to lethal autonomy matters
most in the assessment of military AI featuring fully autonomous lethal
force. Across the broader set of uses, its role is more limited. Against
our expectations, perceived AI risks are positively, not negatively,
associated with support, which suggests that some respondents may
understand military AI risks through a logic of strategic necessity
rather than as a driver of blanket opposition. We also find some
geopolitical variation. Distrust of Russia is associated with greater
support in Europe, while trust in the United States is associated with
greater support in the United Kingdom and the United States itself.

These findings complicate the autonomous-weapons literature. Fully
autonomous lethal AI remains the least accepted use in all nine
countries. But our results suggest that the ``killer robots'' frame
captures only part of the public-opinion environment around military AI.
People also seem to draw on more general views about AI and on broader
orientations toward military force. People's everyday experiences of AI,
seem to matter more than specific military or foreign-policy driven
considerations. Debates over military AI are therefore likely to be
shaped not only by arms-control and autonomous-weapons debates, but also
by everyday encounters with AI and conflicts over civilian AI
governance. Experience with smart toasters might turn out to matter more
than fears of killer robots.

We develop our theoretical framework in the next section. We then
describe the survey design and the six-item measure of support for
military AI. In the empirical section, we first report descriptive
patterns within and across countries, then present multilevel models of
overall associations and country-specific geopolitical patterns. In the
conclusion we present implications for research on public opinion,
autonomous weapons, and civilian AI governance.

\section{How People View Military AI}\label{how-people-view-military-ai}

Most people encounter military AI only indirectly. They have little
direct experience with military decision-making or military technology,
and are unlikely to evaluate specific systems in technical detail.
Instead, they are likely to rely on cognitive shortcuts and accessible
beliefs \autocite{Lau:2001aa,Lupia:1998aa,Zaller:1992aa}. Comparable
patterns of attitude formation are well known in various foreign policy
areas
\autocite{Feldman:1988ab,Hurwitz:1987aa,Kertzer:2017aa,Kertzer:2023aa,Nelson:1996aa}.
The literature points to three different predispositions people may fall
back on, when forming opinions on military AI.

First, we argue that people evaluate military AI as a specific case
within a larger family of uses of a general-purpose technology. People
encounter AI daily in their use of AI-enabled systems and applications
as well as through public debates over AI's expected economic and social
impacts. These experiences may spill over into their views of military
AI. People who associate AI with useful innovation may also be more open
to its military use; those who associate it with risk may be more
skeptical.

An alternative explanation is that people view military AI primarily
through the lens of lethal autonomy. The focus of much of public and
academic debate about autonomous weapons systems focuses on whether
machines should be allowed to select and engage targets without
meaningful human control. These concerns might color views of military
AI more generally, especially as other, more differentiated uses, are
not at the fore of public debate. If true, opposition toward machines
being allowed to make life-and-death decisions should predict opposition
toward military AI more generally.

The third explanation treats military AI as an expression of military
power and the readiness to deploy force in international conflicts. If
these general views dominate people's views of military AI, we expect
people who see military force as legitimate or necessary to be more
supportive of military uses of AI. Those who favor restraint or
diplomatic solutions may instead be more concerned about escalation and
the erosion of human control and oppose military AI.

These explanations look for well-developed and accessible
predispositions as explanations of people's views on military AI.
Alternatively, more context-dependent perceptions might also matter.
People might base their views on military AI on how they see the current
state of the world. Unlike general foreign-policy orientations, these
perceptions can shift in response to wars and crises. People who see the
international order as stable and functioning might feel less of a need
for military AI than those who see the international order as weak and
unable to prevent conflicts. Trust in major powers may work similarly.
If people trust major international powers, military AI might seem as
optional. If they distrust major international powers, military AI might
feel more of a necessity as a response to external threats.

Our argument implies different sources of potential public constraint,
depending on the relative influence of these explanations. If military
AI is evaluated primarily as AI, attitudes toward it will be influenced
by the broader politics of everyday AI. If it is evaluated primarily as
lethal autonomy, then the politics of military AI will be constrained by
moral opposition to machines making life-and-death decisions. If it is
evaluated primarily as military power, then support will depend on
foreign-policy orientations and geopolitical perceptions. Our
theoretical model helps to identify which of these considerations most
strongly organizes people's views of military AI once examined together.

\subsection{Lethality, Automation, and Meaningful Human
Control}\label{lethality-automation-and-meaningful-human-control}

Military AI differs on two dimensions that organize both ethical
critique and operational practice: whether they are directly involved in
the application of lethal force, and how much decision-making authority
they delegate from humans to machines
\autocite{Goldfarb:2022wr,King:2025aa,Scharre:2018aa}. These dimensions
are central to debates over responsibility, accountability, command
authority, and meaningful human control
\autocite{Horowitz:2015aa,Johnson:2024ab,Sparrow:2016aa}.

Lethality captures whether an AI-enabled system is directly involved in
decisions that can cause physical harm or death. Lethal force has a
distinctive moral and legal status; decisions over life and death are
widely understood as requiring human judgment and accountability
\autocite{McMahan:2009aa,Walzer:2015aa}. Ethical critiques of autonomous
weapons focus not only on the consequences of machine error, but also on
the perceived impropriety of delegating lethal judgment to machines
\autocite{Sharkey:2012aa,Sparrow:2007aa}. Legal scholarship on
responsibility gaps reaches a related concern. When harm results from
opaque sociotechnical systems, it may become difficult to identify who
can be held accountable
\autocite{Bo:2022aa,Chengeta:2016aa,Matthias:2004aa,Verdiesen:2021aa}.

Degree of automation captures how far decision-making authority shifts
from human operators to AI-enabled systems. At the lower end, AI-enabled
systems provide options that humans can accept or reject. At the higher
end, AI-enabled systems act without real-time human authorization or
oversight. As automation increases, humans may therefore no longer make
decisions directly, but instead monitor systems that are acting
autonomously. Research on human-machine interaction and automation bias
suggests that this can reduce situational awareness and make it harder
to intervene when needed
\autocite{Endsley:1995aa,Parasuraman:1997aa,Skitka:1999aa}. These
concerns are especially important in military environments, where time
pressure and operational complexity may make human oversight of military
AI more formal than meaningful in practice
\autocite{Cummings:2004aa,Johnson:2024ab}.

This tension points to the importance of meaningful human control. It is
not enough that a human is formally ``in the loop.'' What matters is
whether that person can actually understand what the system is doing,
monitor it in real time, step in when something goes wrong, and be held
responsible for the outcome
\autocite{Chengeta:2017aa,Horowitz:2015aa,Verdiesen:2021aa,Amoroso:2020aa,Boulanin:2020aa}.
The more autonomous the system becomes, the harder this is. This becomes
even more relevant when the system is used in lethal force. High
autonomy combined with lethality is the clearest case in which
meaningful human control matters the most, while potentially being most
fragile.

This leads to our first use-based expectation. Support for military AI
should be lower for lethal than for non-lethal uses and lowest for the
fully autonomous application of lethal force.

\subsection{Three Predispositions}\label{three-predispositions}

Three predispositions are especially likely to structure people's views
of military AI: general orientations toward AI as a technology,
principled opposition to lethal autonomy, and hawk-dove orientations
toward military force. Each links military AI to a different familiar
object. The first treats military AI as an extension of AI in civilian
life. The second treats it as an instance of machines making
life-and-death decisions. The third treats it as a tool of military
power. These explanations are not mutually exclusive. Respondents may
draw on more than one at the same time. The empirical question is
therefore not the search for the one decisive explanation but an
assessment of their relative importance when examined together.

\subsubsection{Military AI as AI: General Benefit and Risk
Perceptions}\label{military-ai-as-ai-general-benefit-and-risk-perceptions}

The first predisposition is toward AI as a general-purpose technology
\autocite{Narayanan:2025aa}. People increasingly encounter AI through
everyday applications and public discourse. These encounters influence
general orientations toward AI that in-turn can travel across domains.
Research on public attitudes toward AI shows that people differentiate
between perceived benefits and perceived risks of AI and that these
perceptions shape support for AI applications across domains
\autocite{Bao:2022aa,Zhang:2019ab}. Studies also suggest that general AI
orientations matter for domain-specific applications, including
public-sector AI, political AI, and national-security contexts
\autocite{Horowitz:2024aa,Jungherr:2024aa,Jungherr:2025aa,Jungherr:2025ac,Kreps:2023aa,Lillemae:2025aa}.
The implication for military AI is straightforward: because most people
know more about AI from everyday contexts than from military ones, they
may extend their general evaluation of AI to military applications.

This does not require respondents to believe that everyday and military
AI are identical. Instead it expects that general AI orientations are
more accessible than detailed knowledge of AI use in the military
domain. If military AI is evaluated primarily as AI, then respondents
who perceive AI as beneficial should support military AI more strongly,
while respondents who perceive AI as risky should support it less
strongly. We therefore expect perceived AI benefits to be positively
associated with support for military AI applications, and perceived AI
risks to be negatively associated with support.

\subsubsection{Military AI as Lethal Autonomy: Principled
Opposition}\label{military-ai-as-lethal-autonomy-principled-opposition}

The second predisposition is people's view of lethal autonomy.
References to military AI may activate concerns about lethal autonomous
weapons. Both the academic literature on autonomous weapons systems and
public activism feature discussions about the moral (im)permissiveness
of having machines autonomously make life-and-death decisions
\autocite{Sharkey:2012aa,Sparrow:2007aa,Carpenter:2020aa,Rosert:2021aa,Bode:2022aa}.
This concern may also come to organize people's views of military AI
generally, independent of the conditions of its specific use.

There are two ways how norms against autonomous weapons can emerge and
matter. One is the discursive development of a norm against autonomous
weapons in international politics and elite discourse. The other focuses
on the emergence of this concern in individual-level attitudes: people
may hold the categorical view that machines should never make
life-and-death decisions, regardless of military usefulness or strategic
context. Our survey measure connects with this second aspect. We measure
whether respondents reject lethal autonomy as morally unacceptable and
examine the connection of this attitude with views on military AI.

Prior work shows that individual-level opposition toward lethal autonomy
is heterogeneous and conditional. Some opposition reflects moral
conviction about removing humans from life-or-death decisions, while
other opposition is conditional on expected outcomes and effectiveness
\autocite{Horowitz:2025aa}. In our study we focus on principled
opposition toward machines making life-and-death decisions without human
control. We do not address outcome-conditional opposition.

If predispositions toward the legitimacy of lethal-autonomy organize
views toward military AI in general, then principled opposition should
reduce support for military AI across all uses. This link should be
strongest where military AI involves lethal force and especially where
lethal force is combined with full autonomy. We therefore expect higher
principled opposition to lethal autonomy to be associated with lower
support for military AI.

\subsubsection{Military AI as Military Power: Hawk-Dove
Orientations}\label{military-ai-as-military-power-hawk-dove-orientations}

The third predisposition connects military AI with attitudes toward
military power more general and the role of force in international
relations. AI-enabled systems are not only technological applications;
they are instruments in the international projection of power.
Accordingly, people may view military AI through broader orientations
toward the use of military force. In foreign-policy research, such
orientations are commonly captured through the distinction between
militant internationalism and cooperative internationalism, often termed
hawk-dove orientations
\autocite{Gravelle:2017aa,Hurwitz:1987aa,Kertzer:2016aa,Kertzer:2023aa,Mader:2023aa}.

For hawks, military strength is central to deterrence and the protection
of national interests
\autocite{Kertzer:2014aa,Rathbun:2016aa,Russett:1990aa}. From this
perspective, military AI is troubling because it may make force easier
to use and harder to assign responsibility when errors occur
\autocite{Horowitz:2018aa,Horowitz:2020aa,Horowitz:2025aa}. Its
operational benefits may therefore be outweighed by the larger political
risks it creates.

In contrast, doves may place less weight on added capability and more on
restraint, legal limits, and the dangers of a new arms race
\autocite{DAgostino:1995aa,Kertzer:2014aa,Rathbun:2016aa}. They may
therefore be more skeptical of military AI because it could lower
thresholds for the use of force, accelerate decision-making in crises,
weaken accountability, or contribute to arms-race dynamics
\autocite{Altmann:2017aa,Asaro:2020aa,Boulanin:2020aa}. For doves, the
potential operational benefits of AI may be outweighed by accidental
escalation and the normalization of automated force.

The hawk--dove provides an explanation of people's view of AI that is
independent of the workings of the underlying technology and associated
discursive postures. Someone who is skeptical of AI in everyday life may
still support its military use if they see it as necessary for national
defense. Conversely, support for AI in civilian settings does not
necessarily imply support for military applications. Respondents may
reject military AI because they oppose the use of force or because they
give priority to restraint. We therefore expect more hawkish respondents
to be more supportive of military AI applications.

\subsection{Contextual Perceptions: International Order and Major-Power
Trust}\label{contextual-perceptions-international-order-and-major-power-trust}

Up until now, we have focused on broad predispositions as possible
organizational principles of people's views of military AI. But these
views are likely also connected to how people view the current
international environment. These contextual perceptions are different
from general hawk-dove orientations about the appropriateness of force.
Contextual perceptions are judgments about the international system or
specific major powers at a given moment.

This reasoning follows public-opinion research in foreign policy, which
separates relatively stable predispositions from more object-specific
perceptions of international politics
\autocite{Kertzer:2017aa,Kertzer:2023aa,Rathbun:2016aa}. What
distinguishes the two is temporal anchoring rather than short-term
volatility. A judgment about the international order, or about a
particular major power, can itself be quite stable; the point is that
its object is the current environment, not a deeper value commitment. We
care about this aspect here because military AI may look more or less
necessary depending on whether people see the world as broadly
rule-governed or as competitive and threatening.

\subsubsection{Confidence in the International
Order}\label{confidence-in-the-international-order}

One contextual perception is confidence in the international rules-based
order. Respondents who think international institutions successfully
secure peace and prosperity tend to view the international environment
as comparatively manageable \autocite{Axelrod:1985aa,Keohane:1995aa}.
From this perspective, military AI may not feel like an especially
urgent or useful innovation. Skeptics reason differently. If current
institutions look ineffective, military-technological advantage becomes
more important
\autocite{Byers:2017aa,Mearsheimer:1990aa,Mearsheimer:1995aa}. In this
context, AI-enabled systems begin to look less like optional innovations
and more like tools needed for the successful projection of power and
maintenance of relative advantage. This is consistent with research
showing that higher threat perceptions increase support for using force,
defense spending, and military innovation more generally
\autocite{DiGiuseppe:2024aa,Horowitz:2016aa,Press:2013aa}. We therefore
expect lower confidence in the international rules-based order, or a
higher sense that the order is dysfunctional, to be associated with
greater support for military AI applications.

\subsubsection{Trust in Major Powers}\label{trust-in-major-powers}

A second contextual factor is trust in major powers. Respondents may see
military AI connected to their views of major powers such as China,
Russia, or the United States. If these actors are seen as responsible,
the international environment may appear less threatening, and the need
for potentially destabilizing military technologies may seem lower. If
they are seen as unreliable or threatening, military AI may instead
appear more necessary as a countermeasure.

We do not expect the views of specific major powers to carry the same
association with support for military AI across countries. Major
international powers like China, Russia, or the United States are seen
differently, depending on their specific posture toward countries.
Distrust of Russia is likely especially salient in European publics
exposed to the post-2022 security shock and may be linked with support
of military AI. Trust in the United States may matter more in countries
where the U.S. security relationship is seen as vital and stable. Trust
in China may matter differently in countries exposed to China as a
strategic competitor, economic partner, or neighboring power. We
therefore expect not only that lower trust in major powers is associated
with higher support for military AI, but also that specific trust
judgments should matter most where the corresponding security
relationship is politically salient.

\subsection{Cross-National Expectations: Shared Structure and Selective
Activation}\label{cross-national-expectations-shared-structure-and-selective-activation}

This theory has implications for cross-national comparison. The
underlying mechanism is general and we do not expect each country to
have a distinct logic of military AI attitudes. Under conditions of
limited information, people rely on broader attitudes and accessible
perceptions to evaluate unfamiliar military technologies. Because this
mechanism is not country-specific, the same broad predictors should
matter across countries. Differences in aggregate attitudes toward
military AI between countries should be driven by different
distributions of attitude holders, not by different attitude structures.
Differences should be compositional, not structural.

The theory also allows for selective activation. Some predictors may
translate into support differently depending on national context. This
should be especially true for geopolitical trust. Trust in Russia, the
United States, or China is not politically equivalent across countries.
Distrust of Russia may carry stronger implications in European publics
where Russia is viewed as a direct security threat. Trust in the United
States may matter more where the U.S. security relationship is central
to national defense or national identity. Trust in China may matter more
where China is a proximate strategic actor. In these cases, the shared
structure is the potential for geopolitical considerations to matter,
the cross-national variation lies in which geopolitical perception
carries meaning depending on the context in which people are embedded.

This leads to two comparative expectations. First, we expect a largely
shared individual-level structure: general AI orientations, principled
opposition to lethal autonomy, hawk-dove orientations, and contextual
perceptions should be relevant across countries. Second, we expect
selective country-specific variation where contextual perceptions,
especially trust in major powers, are tied to nationally salient
security relationships.

This distinction also clarifies what kind of public-opinion constraint
our study can observe. We measure the structure of mass attitudes toward
military AI across nine countries. We do not directly observe how
policymakers, military organizations, technology firms, or civil-society
actors translate those attitudes into policy. Nevertheless, the
structure of public opinion matters because it defines the environment
that actors operate in. If support for military AI is shaped primarily
by general AI orientations, then the political opportunity space for
military AI will move with the broader politics of everyday AI. If
support is shaped primarily by principled opposition to lethal autonomy,
then debates over autonomous weapons and meaningful human control norms
will provide the central public constraint. If support is shaped
primarily by hawk-dove orientations and geopolitical perceptions, then
military AI will follow views on the politics of force and the specific
constellations of geopolitical environments.

Our empirical analysis directly compares these possibilities. We
estimate which predictors are most strongly associated with support for
military AI, how much they account for differences across countries, and
where country-specific patterns remain after these common factors are
taken into account.

\section{Evidence from Nine
Countries}\label{evidence-from-nine-countries}

Our data are part of the \emph{bidt digitalbarometer.international
2026}, a cross-national online survey (n=9,000) that \emph{Ipsos}
fielded for us in January 2026. The survey covers views on digital
technology and Artificial Intelligence in various domains. We developed
a module on military uses of AI for the survey and added measures for
general attitudes toward AI, opposition to lethal autonomy, hawk--dove
orientations, confidence in the international order, and trust in major
powers. We preregistered the study before the start of fieldwork and got
approval by the IRB of the University of Bamberg.

Our analysis focuses on two questions: First, do people distinguish
between military AI applications that differ in lethality and
automation? Second, which broader predispositions and contextual
perceptions explain support for military AI and how does this vary once
we account for the features of the application and differences between
countries?

\subsection{Country Selection}\label{country-selection}

The survey was fielded in China, Finland, France, Germany, Italy, Spain,
Taiwan, the United Kingdom, and the United States, with 1,000
respondents in each country. We selected these countries because they
differ in ways that are relevant to our argument: their place in the
global AI ecosystem, their security environment, and their relationship
to major geopolitical actors.

The first source of variation concerns AI. The United States and China
are central sites of frontier AI development, while Taiwan plays a
strategically important role through its semiconductor industry. The
European countries are also technologically advanced, but their exposure
to AI is shaped partly by technologies, platforms, and governance
debates that often originate elsewhere. General views of AI may
therefore travel into military contexts across all countries, but not
necessarily in the same way.

The second source of variation concerns security context. Finland and
Taiwan face direct security concerns linked to nearby revisionist
powers. In the European cases, Russia's full-scale invasion of Ukraine
has made NATO, European defense, and support for Ukraine more
politically salient. In the United Kingdom and the United States,
debates over alliance politics, the U.S. security role, and military
innovation have a different meaning. China adds a non-democratic
major-power case outside the otherwise largely democratic sample.

This also relates to positions within the international order and
relative exposure to major powers. The European cases are embedded in
NATO, the EU, or both. The United States anchors the transatlantic
alliance system. Finland is distinctive because NATO membership is
recent and alliance politics remains especially salient. Taiwan and
China stand outside these institutions, but both are central to security
politics in the Indo-Pacific. We therefore expect trust in Russia, the
United States, or China to matter most where that relationship is part
of the country's immediate security environment.

The sample also has clear limits. It covers nine publics with developed
AI exposure and substantial geopolitical relevance, but it is not a map
of global public opinion on military AI. Publics in active conflict
zones, including Ukraine, are not included. The same is true for several
states central to military AI development or regional military
innovation, including Russia, Israel, India, Japan, South Korea, and
Turkey. The analysis should therefore be read as evidence about the
structure of military-AI attitudes under these scope conditions. The
questionnaire was fielded in the respective local languages, translated
from a German-language master; details of the translation procedure are
documented in SI B.1.

\subsection{Measures}\label{measures}

We measured support for military AI and all explanatory variables,
unless otherwise noted, using items measured on 7-point scales. Higher
values indicate stronger agreement or greater support (full question
wording and item-level descriptive statistics, along with scale
reliability, are reported in SI B).

\subsubsection{Support for Military AI
Applications}\label{support-for-military-ai-applications}

The dependent variable is a six-item battery on how appropriate
respondents find six different military uses of AI. The items are
arranged on a grid. One axis is lethality, that is, whether the
application directly involves the use of lethal force. The other axis is
automation, meaning how much of the decision is left to the AI system
rather than kept with a human operator. Respondents evaluated each
application on a scale from ``not at all appropriate'' to ``completely
appropriate.''

The six applications are:

\begin{itemize}
\tightlist
\item
  The military uses AI to analyze satellite images and identify enemy
  supply routes.
\item
  AI systems analyze current combat information and recommend safe
  evacuation routes.
\item
  An AI system controls drones that specifically disrupt enemy
  communications.
\item
  AI is used to identify potential targets for future air strikes.
\item
  AI systems make suggestions for attack options based on real-time data
  during ongoing combat.
\item
  An AI-controlled defense system automatically detects and destroys
  enemy units on the battlefield without human intervention.
\end{itemize}

The items form a 2 x 3 design: non-lethal decision support, non-lethal
human-in-the-loop use, non-lethal autonomous use, lethal decision
support, lethal human-in-the-loop use, and fully autonomous lethal
force. This structure allows us to examine whether support changes
systematically as applications become more lethal and more automated.

The items also span major operational categories identified in standard
taxonomies of military AI: intelligence and surveillance, logistics and
operational support, electronic warfare, targeting, command-and-control
decision support, and autonomous lethal systems
\autocite{Horowitz:2018aa,Scharre:2018aa}. The battery is not intended
to exhaust the full universe of military AI applications. Rather, it
samples applications that vary meaningfully along the ethical and
political dimensions central to our theory: involvement in lethal force
and delegation of decision-making authority.

We preregistered a single six-item index as our primary dependent
variable. As preregistered, we also conducted factor analyses to examine
the structure of the six-item battery. A two-factor solution separating
lethal and non-lethal applications fits only marginally better than a
single-factor solution, and the two factors are highly correlated (see
SI C.1 for the complete analysis). The fully autonomous lethal item
contributes most strongly to the second factor. Cross-country
measurement-invariance tests support configural and metric invariance
for both one- and two-factor solutions, allowing correlational analysis
(see SI C.2). Scalar invariance, which is required for direct mean
comparisons, holds only for the two-factor solution. We therefore use
all six application evaluations as a single index in the explanatory
models as preregistered and compare lethal and non-lethal dimensions
separately in the descriptive analysis.

\subsubsection{General AI Orientations}\label{general-ai-orientations}

We measure general orientations toward AI with two four-item batteries
capturing perceived societal benefits and perceived societal risks. The
batteries were developed and tested in prior research on public
attitudes toward AI \autocite{Jungherr:2024aa,Jungherr:2025aa}. They are
measured independently because perceived benefits and perceived risks
need not be opposite ends of a single scale. Respondents may see AI as
both highly beneficial and highly risky, as neither beneficial nor
risky, or as primarily one or the other.

The AI-benefits battery captures the extent to which respondents view AI
as a technology with positive societal consequences. Respondents
evaluated whether AI will drive significant economic expansion in their
country, help governments plan for the future and manage crises more
efficiently, help parties communicate more successfully with voters, and
help humanity address existential threats more successfully. Higher
values indicate stronger perceived societal benefits of AI.

The AI-risks battery captures the extent to which respondents associate
AI with social, economic, political, and existential dangers.
Respondents evaluated whether AI is likely to cause widespread job
displacement and unemployment, whether increasing AI decision-making
risks loss of control over people's lives, whether AI will allow parties
to manipulate voters more successfully, and whether unchecked AI
development could pose existential threats to humanity. Higher values
indicate stronger perceived societal risks of AI.

The two batteries allow us to test whether support for military AI
follows general evaluations of AI as a technology. If military AI is
evaluated primarily as an extension of AI in civilian life, perceived AI
benefits should be associated with higher support for military AI, while
perceived AI risks should be associated with lower support.

\subsubsection{Principled Opposition to Lethal
Autonomy}\label{principled-opposition-to-lethal-autonomy}

We measure principled opposition to lethal autonomy with a three-item
battery developed from recurring themes in the autonomous-weapons
literature and public debates over meaningful human control. Respondents
indicated agreement with three statements: that machines should never be
allowed to make decisions about life and death without human control;
that autonomous weapons violate fundamental ethical norms; and that the
use of lethal autonomous weapon systems is morally unacceptable.

The measure is designed to capture a specific subtype of
autonomous-weapons opposition: principled rejection of machines making
life-and-death decisions. It therefore captures opposition grounded in
moral principle rather than opposition that depends on expected
consequences, such as battlefield effectiveness, civilian protection,
compliance with international humanitarian law, escalation risk, or
adversary behavior. Higher values indicate stronger principled
opposition to lethal autonomy.

This distinction is important for interpreting the results. The battery
is well suited to testing whether a moral opposition to lethal autonomy
generalizes across a broader set of military AI applications. It is less
suited to testing outcome-conditional forms of autonomous-weapons
skepticism. Those would require items that explicitly vary expected
performance, legality, civilian harm, or strategic consequences.

\subsubsection{Hawk-Dove Orientations}\label{hawk-dove-orientations}

We measure orientations toward the use of military force with a
four-item battery adapted from prior research \autocite{Mader:2025aa}.
Respondents evaluated whether war is sometimes necessary to protect the
interests of their own country; whether their country needs a strong
military in order to achieve anything in international politics; whether
their country should do everything in its power, including the use of
military force, to prevent attacks by expansionist states; and whether
the use of military force is never justified. The final item is reverse
coded, so that higher values indicate more hawkish orientations.

This measure captures a general disposition toward military force rather
than a specific attitude toward AI. It allows us to test whether support
for military AI is influenced by the broader politics of force,
deterrence, and strategic competition. If respondents evaluate military
AI as an instrument of military power, more hawkish respondents should
express greater support for military AI applications.

\subsubsection{Contextual Perceptions of the International
Environment}\label{contextual-perceptions-of-the-international-environment}

We measure contextual perceptions with two batteries: confidence in the
international rules-based order and trust in major powers. These
variables capture respondents' assessments of the current international
environment rather than general orientations toward military force.

\paragraph{Confidence in the International
Order}\label{confidence-in-the-international-order-1}

Confidence in the international order is measured with three items.
Respondents evaluated whether international rules and cooperation help
secure peace, whether common rules between countries have contributed to
greater prosperity worldwide, and whether the international order is
capable of effectively managing global crises.

This measure captures whether respondents see the international
environment as rule-governed and manageable or as weakly
institutionalized and crisis-prone. We expect lower confidence in the
international order to be associated with greater support for military
AI, because AI-enabled military capabilities may appear more necessary
when international institutions are seen as unable to manage conflict
and constrain state behavior.\footnote{H5a was preregistered in terms of
  perceived dysfunction, while the variable itself was preregistered as
  ``Confidence in international order.'' We use the preregistered
  confidence measure, so the H5a prediction translates into a negative
  coefficient.}

\paragraph{Trust in Major Powers}\label{trust-in-major-powers-1}

Trust in major powers is measured by asking respondents how much they
trust China, Russia, and the United States to behave responsibly in the
world. Respondents evaluated each country on the same scale. In the
preregistered models, we use an index of trust in major powers. In an
exploratory analysis, we disaggregate the index into trust in the United
States, trust in China, and trust in Russia.

The aggregate trust measure captures generalized confidence that major
powers behave responsibly in world politics. Lower trust indicates a
more threatening perceived international environment. Although a
deviation from our preregistration, the disaggregated measures allow us
to test whether specific geopolitical relationships matter differently
across countries. For example, distrust of Russia may carry different
implications in European publics than in Taiwan, China, or the United
States; trust in the United States may be especially relevant in publics
where the U.S. security relationship is politically central.

\subsection{Analytical Approach}\label{analytical-approach}

We proceed in four steps. First, we present descriptive results for the
six military AI applications. This analysis examines whether publics
distinguish between lethal and non-lethal applications. It shows whether
fully autonomous lethal force occupies a distinctive position in the
attitude structure.

Second, to test our central hypotheses about the three attitude anchors,
we estimate Bayesian multilevel linear regression models predicting
support for military AI. We use multilevel models with country-level
varying intercepts to account for baseline differences in support across
publics. Individual-level predictors include perceived AI benefits,
perceived AI risks, principled opposition to lethal autonomy, hawk-dove
orientations, confidence in the international order, trust in major
powers, and demographic controls for age, gender, and education. As
defined in the preregistration, the models are estimated on multiple
imputed data using 20 imputed datasets and the full sample of 9,000
respondents to address non-responses (people selecting ``cannot
assess'') for the items of the outcome variable (for more information
about non-responses and the imputation procedure, see SI C.3 and D).

Third, we compare the relative predictive weight of general AI
orientations and principled opposition to lethal autonomy using
standardized coefficients. This comparison directly tests whether
military AI attitudes are better explained by general views of AI as a
technology or by principled opposition to machines making life-and-death
decisions.

Fourth, we examine cross-national structure. We estimate how much the
between-country variance is reduced by the inclusion of individual-level
predictors and compare fixed-slope and varying-slope specifications with
a baseline model that includes only varying intercepts and no
individual-level predictors, using leave-one-out cross-validation (LOO)
\autocite{Vehtari:2017aa}. This allows us to distinguish compositional
cross-national variation (i.e.~differences arising because publics
differ in average levels of AI optimism, hawkishness, or threat
perceptions) from selective structural variation, where the same
predictor matters differently across countries. As an exploratory
analysis not specified in the preregistration, we disaggregate trust in
major powers into trust in the United States, China, and Russia and
estimate country-specific slopes for each trust item.

All main-text analyses use preregistered model specifications unless
otherwise noted. Robustness checks using alternative inclusion criteria
and complete-case specifications are reported in SI D.4.

\section{The Structure of Military AI
Attitudes}\label{the-structure-of-military-ai-attitudes}

We begin by examining the descriptive structure of support across the
six military AI applications. We first examine how people think about
military AI and whether they differentiate between different types of
military AI, and if so which. We then identify attitude structures that
explain people's support or opposition to military AI and whether these
structures vary between countries.

\subsection{Lethality and the Limits of
Automation}\label{lethality-and-the-limits-of-automation}

\begin{figure}[htbp]
  \centering
  \includegraphics[width=0.95\textwidth]{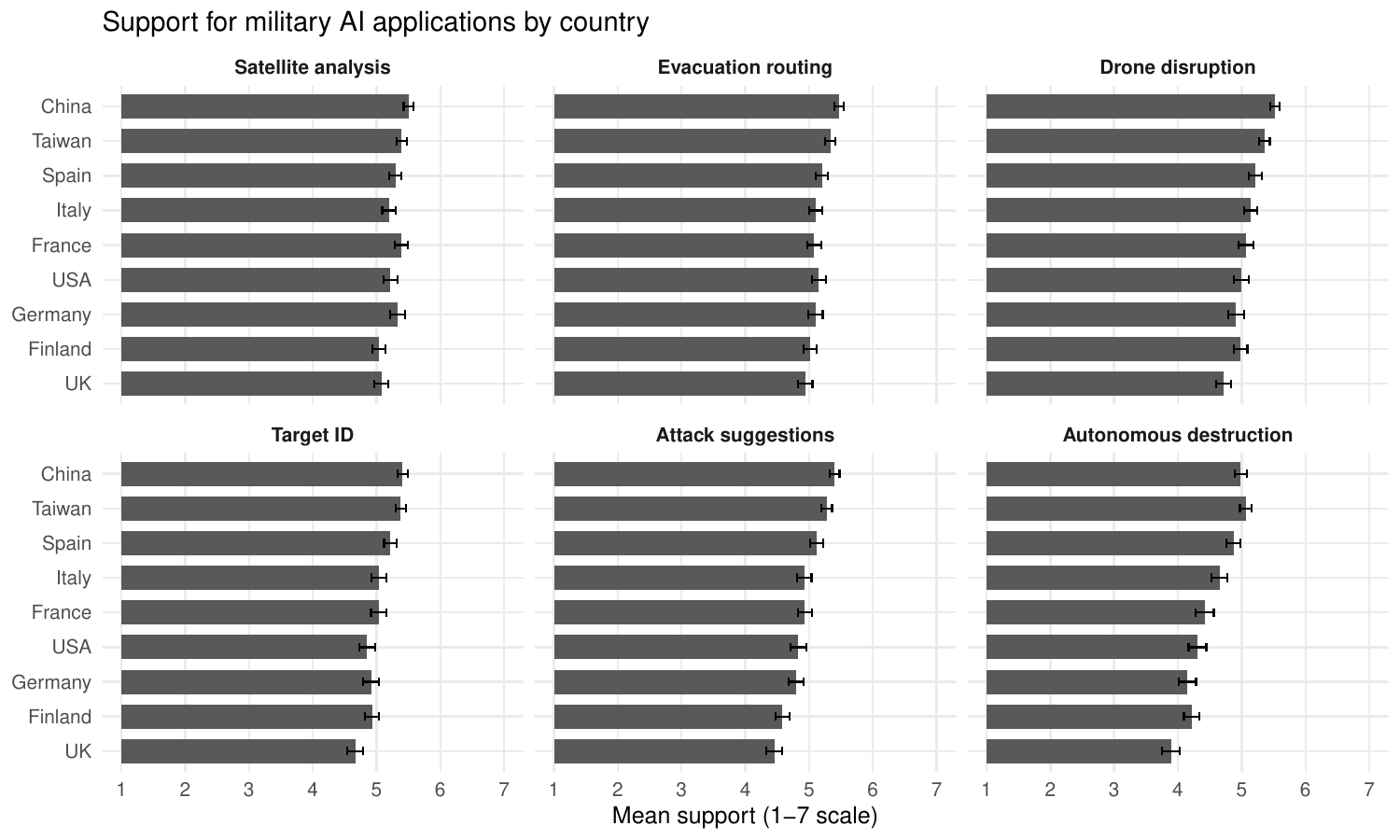}
  \caption{Mean support for different military uses of AI. 95\% CIs
shown. Countries ordered by the overall mean for the six-item
military-use-of-AI index. The bottom row shows AI uses with higher
lethality.}
  \label{fig:1}
\end{figure}

Across all six applications, mean support in every country sits above
the scale midpoint of 3.5 (Figure~\ref{fig:1}). Publics in our sample were
therefore not, in the aggregate, generally opposed to military uses of
AI. They tilted toward qualified acceptance across the application
space. This does not mean that all applications were evaluated equally.
The pattern is conditional: support depends on what military AI is used
for and how much authority is delegated to the system.

\begin{figure}[htbp]
  \centering
  \includegraphics[width=0.95\textwidth]{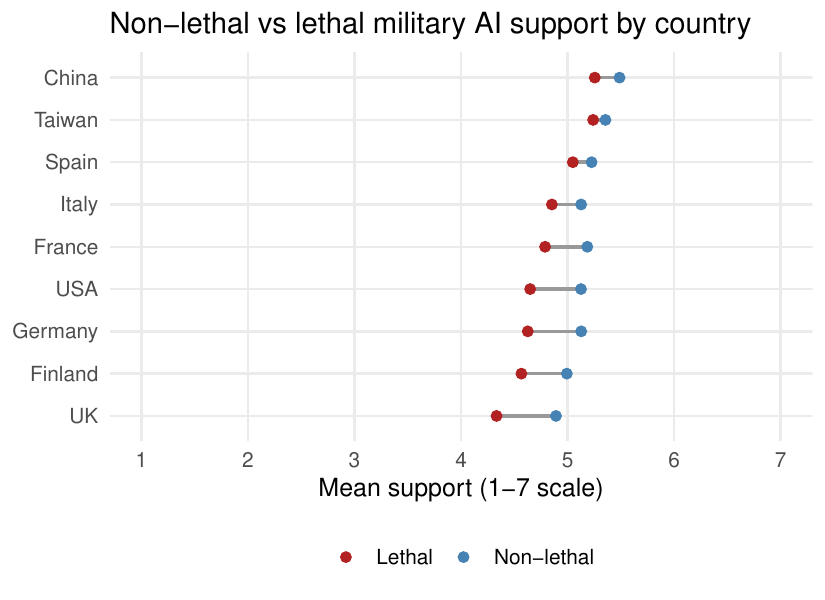}
  \caption{Mean comparison between lethal and non-lethal military
uses of AI. Means are calculated on all available responses before
imputation. Respondents contributed to an index if they answered at
least one item. All differences are significant after Bonferroni
correction for nine tests.}
  \label{fig:2}
\end{figure}

Two descriptive patterns stand out. First, lethality is a strong
organizing dimension. In every country, mean support for the three
non-lethal applications exceeds mean support for the three lethal
applications (Figure~\ref{fig:2}). These differences remain significant even after
Bonferroni correction for all nine country-level paired t-tests. The gap
is especially pronounced in the European samples and smaller in China
and Taiwan, where baseline support is higher across the item battery.
This pattern suggests that the public responds to the moral dimension of
lethal force even when they do not reject military AI in general.

The descriptive results also show that support declines as automation
increases. Within the lethal dimension, decision-support applications
receive the highest support, human-in-the-loop applications fall in the
middle, and fully autonomous applications receive the lowest support.
The three non-lethal applications cluster more closely together. The
drop from human-in-the-loop to fully autonomous use is larger for lethal
than for non-lethal applications. Removing real-time human oversight,
therefore, matters especially when AI is linked to lethal force.

The fully autonomous lethal application is the least supported item in
every country. This provides the clearest descriptive evidence that
meaningful human control matters. Respondents do not simply reject
automation, nor do they distinguish only between lethal and non-lethal
applications. Support falls most sharply when both elements are present:
lethal force and the absence of human oversight.

Still, the autonomous-lethal item falls below the scale midpoint only in
the United Kingdom (M = 3.89), and support for other military AI
applications remains comparatively high. Public opinion, then, is not
simply opposed to military AI. Respondents are most accepting of
non-lethal and human-supervised uses, but support declines when AI is
used to exercise fully autonomous lethal force.

\subsection{What Explains Support?}\label{what-explains-support}

\begin{figure}[htbp]
  \centering
  \includegraphics[width=0.95\textwidth]{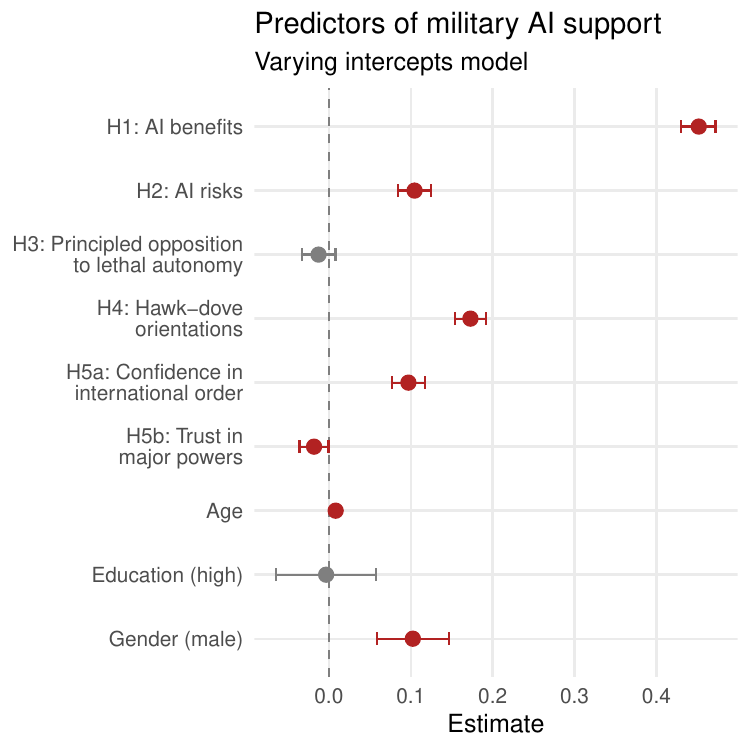}
  \caption{Forest plot of unstandardized coefficients from a
Bayesian multilevel linear regression with varying intercepts for
countries predicting support for military AI applications, estimated on
multiple imputed data (m = 20, N = 9,000). Estimates with 95\% CIs.
Predictors whose CIs exclude zero are highlighted in red.}
  \label{fig:3}
\end{figure}

We next estimate Bayesian multilevel linear regression models predicting
support for military AI. The models include country-level intercepts and
the individual-level predictors. The unstandardized coefficients
reported in Figure~\ref{fig:3} indicate the change in support on the 1--7 scale
associated with a one-unit increase in each predictor. The models are
estimated on 20 imputed datasets (N = 9,000).

Our analysis shows that perceived AI benefits are the largest predictor
in the model. Respondents who view AI as beneficial for society are
substantially more supportive of military AI applications (b = 0.45,
95\% CI {[}0.43, 0.47{]}). This finding is consistent with the argument
that general orientations toward AI travel into evaluations of military
AI. For many respondents, military AI appears to be evaluated partly
through the same broader technology orientation that shapes attitudes
toward AI in everyday contexts.

Hawk-dove orientations are the second major predictor. More hawkish
respondents are more supportive of military AI (b = 0.17, 95\% CI
{[}0.15, 0.19{]}). This association indicates that military AI is also
evaluated as a capability embedded in the broader politics of force.
Respondents more accepting of military force appear more willing to
accept AI-enabled military capabilities.

The model provides no support for the principled opposition to lethal
autonomy when support for military use of AI is measured as a
one-dimensional index. Principled opposition to lethal autonomy is not
credibly associated with overall support for military AI (b = -0.01,
95\% CI {[}-0.03, 0.01{]}). This does not mean that lethal autonomy is
irrelevant to public opinion. The descriptive results show that fully
autonomous lethal force is the least-supported application in every
country. Rather, the model indicates that principled opposition to
lethal autonomy does not generalize strongly across the broader battery,
which includes non-lethal, human-supervised, and decision-support
applications.

The contextual predictors show a more limited but still informative
pattern. Higher confidence in the international rules-based order is
associated with greater support for military AI (b = 0.10, 95\% CI
{[}0.08, 0.12{]}). This contrasts with our expectation that respondents
who see the international environment as less manageable and less
rule-governed may view AI-enabled military capabilities as more
necessary. Instead, respondents who express stronger confidence in
international rules and institutions are more supportive of military AI
applications. Furthermore, pooled trust in major powers is only
borderline credibly associated with support (b = -0.02, 95\% CI
{[}-0.04, -0.00{]}). As we show below, this aggregate masks
country-specific patterns tied to particular geopolitical relationships.

Perceived AI risks are positively, not negatively, associated with
support for military AI (b = 0.10, 95\% CI {[}0.08, 0.12{]}). This
finding runs counter to our preregistered expectation. However, the
effect is smaller than the effect of perceived AI benefits. We return to
this result in the interpretation section. For now, the important point
is that general AI-risk perceptions do not translate straightforwardly
into opposition to military AI.

The analysis shows that support for military AI is structured most
strongly by general AI-benefit perceptions and hawk-dove orientations.
The strongest sources of support are therefore not located exclusively
in the autonomous-weapons frame. Instead, military AI attitudes appear
to draw on broader views of AI as a technology and the international
projection of force.

\subsection{General AI Orientations and Principled
Opposition}\label{general-ai-orientations-and-principled-opposition}

To compare the relative predictive weight of general technology
orientations and principled opposition to lethal autonomy, we used the
\emph{hypothesis} function in \emph{brms} to compare standardized effect
sizes \autocite{Burkner:2017aa}. Standardized coefficients allow direct
comparison across predictors measured on different scales.

The comparison reinforces the pattern from the unstandardized model. The
association of AI-benefit perceptions with support for military AI is
large ($\beta$ = 0.63, 95\% CI {[}0.60, 0.66{]}). By contrast, the
standardized association of principled opposition to lethal autonomy is
close to zero ($\beta$ = -0.02, 95\% CI {[}-0.04, 0.01{]}). The difference
between the two associations is decisive (Δ$\beta$ = -0.61, 90\% CI {[}-0.64,
-0.58{]}, BF \textgreater{} 1000).

Perceived AI risks also have a stronger association with military AI
support than principled opposition does, although the effect is much
smaller than that of AI-benefit perceptions ($\beta$ = 0.13, 95\% CI {[}0.10,
0.15{]}; Δ$\beta$ = -0.11, 90\% CI {[}-0.13, -0.09{]}, BF \textgreater{}
1000). The direction of this association is positive, again indicating
that risk perceptions do not operate as simple opposition to military
AI.

These comparisons sharpen our core empirical claim. Across the broad
six-item index, attitudes toward military AI are more strongly predicted
by general orientations toward AI as a technology than by principled
opposition to lethal autonomy. This does not imply that principled
opposition is irrelevant for fully autonomous lethal weapons
specifically. Nor does it imply that people are indifferent to the need
for meaningful human control. Rather, it shows that the
autonomous-weapons frame does not dominate evaluations of military AI
once the outcome includes a wider range of military applications.

In a non-preregistered follow-up analysis, we ask whether the weak
association between principled opposition and the broad military AI
index masks stronger connections with military use cases more directly
connected to the underlying autonomous weapons systems (AWS) debate. We
re-estimated the standardized model using the fully autonomous lethal
item as the dependent variable (see SI D.6). Principled opposition
emerges as a substantial negative predictor ($\beta$ = -0.17, 95\% CI
{[}-0.21, -0.13{]}) roughly eight times its association with the broader
index and statistically indistinguishable in absolute magnitude from
AI-risk perceptions ($\beta$ = 0.19, 95\% CI {[}0.15, 0.23{]}). AI-benefit
perception remains the strongest predictor ($\beta$ = 0.74, 95\% CI {[}0.69,
0.78{]}), but its dominance over principled opposition shrinks from
approximately thirty-fold in the index model to roughly four-fold for
the autonomous-lethal item. The AWS-frame anchor is therefore not absent
from public reasoning about military AI; rather, it is concentrated in
the application that most directly corresponds to its conceptual target.
The same gradient is visible in models estimated separately for the
lethal and non-lethal sub-indices (see SI D.7), where principled
opposition shifts from a small positive association on the non-lethal
sub-index ($\beta$ = 0.04, 95\% CI {[}0.01, 0.07{]}) to a negative association
on the lethal sub-index ($\beta$ = -0.08, 95\% CI {[}-0.10, -0.05{]}).

This finding should be read against the operationalization of the
autonomous-lethal item, which describes a defensive system rather than
the offensive autonomous weapons scenario most central to the AWS-norm
literature. Even with this more conservative framing, the item is the
least supported in every country, and principled opposition is a
substantial predictor of opposition to its use. An offensive
operationalization would therefore likely yield an even stronger
association with principled opposition.

\subsection{Compositional Variation and Selective
Activation}\label{compositional-variation-and-selective-activation}

The final step examines how the structure of support varies across
countries. The baseline model, with varying intercepts but no
individual-level predictors yields an intraclass correlation coefficient
(ICC) of 0.048. Approximately five percent of the total variance in
support for military AI lies between countries. Most variation is
therefore within countries rather than between them. Despite differences
in security environment, military posture, regime type, and AI ecosystem
across the nine countries, respondents within each country differ from
one another much more than the average respondent in one country differs
from the average respondent in another.

Adding individual-level predictors reduces country-level variance by 48
percent ($\tau_{00}$ = 0.079 to 0.041). This indicates that a meaningful share
of cross-national variation is compositional: countries differ partly
because their publics differ in average levels of AI optimism,
hawkishness, confidence in the international-order, and related
attitudes. The result supports our article's expectation that
cross-national differences in military AI support arise not primarily
from wholly distinct national attitude structures, but from the
distribution of broadly shared predictors.

\begin{figure}[htbp]
  \centering
  \includegraphics[width=0.95\textwidth]{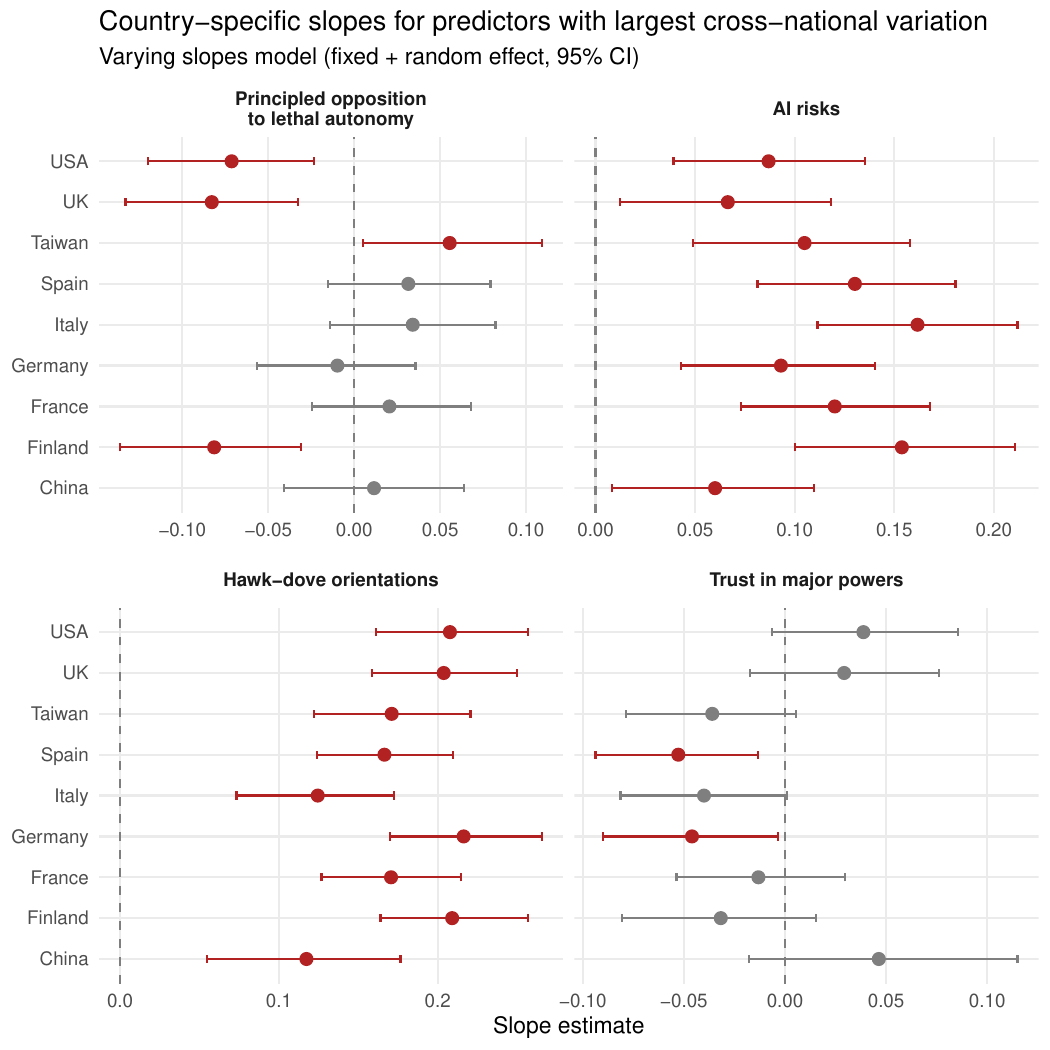}
  \caption{Country-specific slopes from a Bayesian multilevel
linear regression with varying intercepts and slopes, estimated on
multiple imputed data (m = 20, N = 9,000). Panels show the four
predictors with the largest random-slope standard deviations. Estimates
for slopes (fixed effect + random effect) with 95\% CIs. Predictors with
CIs that exclude zero are highlighted in red.}
  \label{fig:4}
\end{figure}

At the same time, the predictors do not operate identically everywhere.
A comparison using LOO favors the varying-slopes model over the
fixed-slopes model (ΔELPD = 48.2, SE = 14.8; see SI D.3 for more
details). Some associations therefore vary across countries in
substantively meaningful ways (Figure~\ref{fig:4}). This is consistent with the
expectation of selective activation: certain attitudes and perceptions
become more consequential where they connect to nationally salient
security relationships or political contexts.

The country-specific slopes show that AI-risk perceptions and hawk-dove
orientations have relatively consistent associations across countries.
Trust in major powers varies more. This is theoretically important
because trust in major powers is the contextual perception most likely
to acquire different meanings across national settings. Trust in Russia,
the United States, or China is not politically equivalent in Germany,
Taiwan, China, the United Kingdom, or the United States.

To examine this more directly, we estimate non-preregistered model that
disaggregates the trust-in-major-powers index into trust in the United
States, trust in China, and trust in Russia, with country-specific
slopes for each item. At the pooled level, none of the three trust items
is credibly associated with military AI support: trust in the United
States (b = 0.02, 95\% CI {[}-0.01, 0.04{]}), trust in China (b = -0.01,
95\% CI {[}-0.03, 0.02{]}), and trust in Russia (b = -0.02, 95\% CI
{[}-0.05, 0.02{]}). This mirrors the aggregate weak result of the pooled
trust index.

\begin{figure}[htbp]
  \centering
  \includegraphics[width=0.95\textwidth]{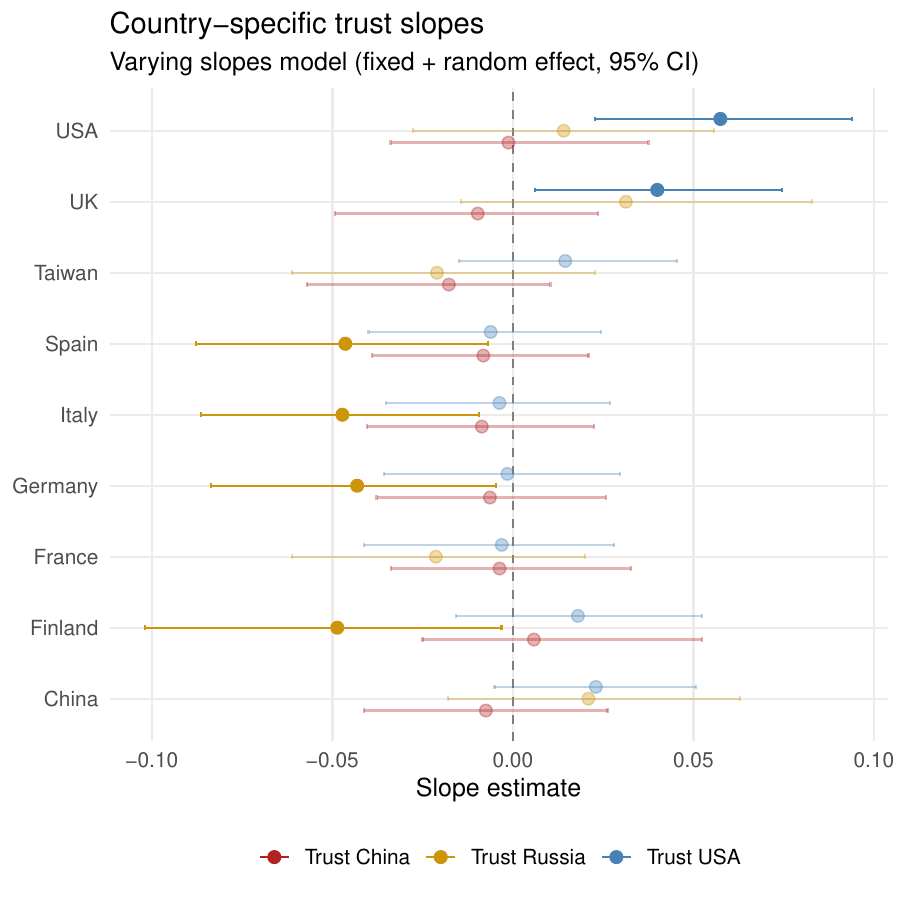}
  \caption{Country-specific trust slopes from a Bayesian multilevel
linear regression with varying intercepts and slopes for disaggregated
trust items, estimated on multiple imputed data (m = 20, N = 9,000).
Estimates for slopes (fixed effect + random effect) with 95\% CIs.
Predictors with CIs that exclude zero are shown at full opacity.}
  \label{fig:5}
\end{figure}

The country-specific slopes differentiating by international power,
however, reveal clear geopolitical patterning (Figure~\ref{fig:5}). Trust in the
United States is positively associated with military AI support in the
United States itself (b = 0.06, 95\% CI {[}0.02, 0.09{]}) and in the
United Kingdom (b = 0.04, 95\% CI {[}0.01, 0.07{]}). Trust in Russia
shows the strongest cross-national variation (SD = 0.04) and is
negatively associated with support in Germany (b = -0.04, 95\% CI
{[}-0.08, -0.01{]}), Italy (b = -0.05, 95\% CI {[}-0.09, -0.01{]}),
Spain (b = -0.05, 95\% CI {[}-0.09, -0.01{]}), and Finland (b = -0.05,
95\% CI {[}-0.10, -0.00{]}). In these countries, higher distrust of
Russia is associated with higher support for military AI. Trust in China
shows little cross-national variation and no country-specific
associations.

We read this as evidence of selective activation. Trust in major powers
does not work as a uniform predictor across the nine countries. It
matters for specific powers in specific countries along preexisting
geopolitical constellations. Distrust of Russia is associated with
support for military AI in several European countries that are dealing
with the impact of Russia's ongoing war of aggression against Ukraine.
Trust in the United States matters in the United States and the United
Kingdom. For US respondents this hints at them seeing AI as an extension
of their country's ability to act on the international stage. For the UK
this indicates that the security relationship with the US is very much
at the heart of the perception of the international environment. Trust
in China does not show clear associations anywhere.

Overall, people draw distinctions between military AI applications.
Support drops sharply once AI is given autonomous control over lethal
force. Across the other applications, what mostly moves support is how
useful people find AI in general and how hawkish they are; principled
opposition to lethal autonomy does much less work than the prior
literature might indicate. Variation across countries is modest. The
clear exception are perceptions of the geopolitical environment in the
form of trust in great powers, which matters mainly where the
relationship with a particular major power is already salient.

\subsection{Interpretation: Risk Perception, Principled Opposition, and
the Conditions of
Constraint}\label{interpretation-risk-perception-principled-opposition-and-the-conditions-of-constraint}

Two findings ask for further interpretation. First, we find AI-risk
perceptions to be positively associated with support for military AI,
contrary to our preregistered expectation. Second, principled opposition
to lethal autonomy is not associated with opposition toward military AI.
These results complicate a simple extension of general AI skepticism or
principled opposition to lethal autonomy, as neither straightforwardly
translates into lower support for military AI overall. The positive
association for AI-risk perceptions is especially striking, suggesting
several possible explanations.

The positive coefficient for AI risks may reflect a security dilemma
logic. Some respondents may worry about AI precisely because they see it
as powerful. But if they expect that other states are already developing
military AI, that concern may not lead them to reject it. It may instead
make the technology seem necessary. In this view, a military that avoids
AI could appear unprepared or vulnerable. This fits threat-perception
research, which links perceived competition and adversary behavior to
greater support for autonomous and AI-enabled military capabilities
\autocite{DiGiuseppe:2025aa}.

Another possible interpretation is that the risk item partly captures
salience. Benefits and risks are not always opposites in people's minds.
Some respondents may see AI as equally promising and dangerous. For
them, high-risk perceptions may simply mean that AI is a consequential
technology, not that it should be kept out of public use. This is close
to arguments in the literature on risk-benefit perceptions and the
affect heuristic \autocite{Finucane:2000aa,Slovic:2007aa}. People often
judge complex technologies by their overall orientation toward them,
rather than by cleanly separating advantages from dangers. Survey work
on ambivalent attitudes toward AI suggests the same pattern
\autocite{Bao:2022aa}.

The cross-sectional design of our study cannot distinguish among these
interpretations. Future research could measure AI attentiveness
directly, and more directly test whether strategic cues translate
military AI risks into support for military AI.

We also need to take a closer look at the role of principled opposition
to lethal autonomy. In the single-index models, this measure is not
associated with overall support for military AI. That does not mean
respondents are indifferent to human control. Fully autonomous lethal
force is the least supported application in every country. People
clearly care if AI is described as using lethal force without human
oversight.

We find that opposition to lethal autonomy is not a predictor of
attitudes toward military AI in general (measured by us as a bundle of
six uses ranging from non-lethal decision support to fully autonomous
lethal force) but that it is connected with concerns about one task, the
fully autonomous lethal application. This indicates to us that concerns
about ``killer robots'' are real but limited in their application to
military AI in general. The non-preregistered additional analysis in SI
D.6 supports this reading. When we use the autonomous-lethal item as the
outcome, principled opposition becomes a clearly negative predictor.
This suggests that principled opposition matters for lethal autonomy
proper but not for military AI as a broader category. Principled
opposition to lethal autonomy as a predictor is also narrow in a
different way. It measures a categorical objection to lethal autonomy
but does not speak to more conditional worries or framing effects,
aspects that should be investigated in future work.

The broader conclusion is straightforward. People do not reject military
AI as such. Their unease is clearest when AI is tied to lethal force
without human oversight. For other uses in the military AI index, such
as decision support, intelligence analysis, or logistics, general views
of AI, attitudes toward military force, and perceptions of the
international environment matter more. Public constraint is therefore
likely to be selective rather than general. Thus, it should be strongest
for fully autonomous lethal force and weaker for military AI as a wider
set of tools.

\section{Everyday AI and the Political Opportunity Space for Military
AI}\label{everyday-ai-and-the-political-opportunity-space-for-military-ai}

Military AI is often debated through the frame of the ``killer robot''
that removes humans from life-and-death decisions. Our results show that
this frame matters, but only up to a point. Across nine countries,
support is lowest for fully autonomous lethal force, where public unease
is clearest. For other military uses of AI, attitudes rest less on
meaningful human control and more on general views of AI and hawk-dove
orientations toward military force.

The public-opinion environment for military AI is therefore neither
simply prohibitive nor unconditionally permissive. Mean support is
generally above the midpoint of the scale, including for several lethal
applications, but respondents distinguish among uses. Support declines
with lethality and higher automation, and is lowest when lethal force is
delegated without human oversight.

The strongest predictor in our models is perceived AI benefit.
Respondents who see AI as beneficial for society are more supportive of
military AI across the six-item battery. This finding suggests that
publics often evaluate military AI as an extension of AI as a
general-purpose technology. The same attitudinal foundation that informs
civilian AI politics (i.e., perceived benefits and risks of AI) also
predicts evaluations of military AI. Experiences with smart toasters
might matter more than fears of killer robots for how people think of
military AI. Everyday experiences with AI and exposure to general AI
discourse may turn out to be the dominant force shaping political
opportunities structures for military AI.

This finding matters for both security studies and AI governance. The
two fields often treat military and civilian AI as separate domains. Our
results suggest that both areas are linked, at least where public
opinion is concerned. When people assess military AI, they do not appear
to do so with lethal military application front of mind. Instead, they
seem to view military AI in line with their broader views of AI. Thus,
public discourse surrounding civilian AI, its benefits as well as its
risks, could affect the trust people place in the military's use of AI.

At the same time, the autonomous-weapons literature remains relevant. In
every country, respondents were least supportive of fully autonomous
lethal force. This shows that concern about meaningful human control is
real and not confined to one national context. But our findings also
show limits. Opposition to lethal autonomy does not automatically extend
to military AI more broadly, at least when the issue is not framed
primarily in terms of high lethality and autonomy alone. Whether
autonomous weapons become the dominant frame for military AI politics
may therefore depend on political actors. They would have to connect the
unease around fully autonomous lethal force to other military uses of
AI. Without that connection, many people appear to judge these
applications through broader views of AI and military force.

The second strong predictor is hawk-dove orientation. Respondents who
are more willing to accept military force as an element of international
affairs are also more open to military AI, even after we account for
their general views of AI. The military setting, therefore, matters in
its own right. People who are already comfortable with military power
are more likely to accept AI in that context, while those who are uneasy
about military power are less likely to do so.

Geopolitical perceptions matter, but less uniformly. Pooled trust in
major powers is not a strong predictor of support. Once we separate the
indicators by country, however, a clearer pattern emerges. Distrust of
Russia is associated with greater support for military AI in Finland,
Germany, Italy, and Spain. Trust in the United States is associated with
greater support in the United Kingdom and in the United States itself.
These results suggest that geopolitical trust matters most where the
actor in question is central to how people think about national
security. Military AI attitudes, therefore, have a common structure
across countries, but they are still shaped by specific security
contexts.

The positive association between AI-risk perceptions and military-AI
support is less intuitive. Concern about AI risks does not automatically
mean rejection of military AI. One possibility is a security-dilemma
logic: if AI is dangerous, respondents may think their own state still
needs access to it because adversaries will use it. Another possibility
is familiarity. Respondents who know more about AI may see both its
benefits and its risks. The next step is to empirically separate these
possibilities, especially the roles of military-specific risk
perceptions, AI familiarity, and expectations about adversary
development.

Our study has a number of limitations. First, we use a cross-sectional
design and cannot make causal claims. Furthermore, as noted in our
discussion of case selection, the sample excludes active conflict zones
and several states central to military-AI development, so our findings
should be interpreted with these scope conditions in mind. For most
predictors, variation between the nine countries is limited, and some
findings may generalize beyond these cases, which could be further
confirmed in future studies. We have already discussed further
limitations of our variable operationalizations above, for example, for
autonomous-weapons opposition.

These limitations leave several questions for future research. This
includes experimental tests whether support changes when military AI is
described in more concrete terms, for example, with different levels of
human oversight or different battlefield outcomes. Comparative and
longitudinal work should also examine whether public controversies shift
attitudes over time. Finally, future work should ask whether and under
what conditions policymakers and military organizations actually treat
public opinion as a constraint on military AI.

Our article provides a helpful baseline on international public opinion
on military AI at a point when this technology becomes a common feature
in international conflict and the projection of power. We find a
conditionally permissive consensus of people viewing military AI
somewhat openly, while being more critical of fully autonomous lethal
uses. This is an important corrective to the extent that elite and media
discourse might suggest that public opinion would be generally critical
of military uses of AI. Instead of seeing military AI exclusively
through the frame of ``killer robots'', people seem to treat it largely
as an extension of their views about AI more generally. This indicates
that the political opportunity space for the development and deployment
of military AI is not primarily formed in specialist discourse, but is
shaped by the larger societal climate of openness or skepticism toward
AI as a technology.

\section{Acknowledgements}\label{acknowledgements}

The authors used ChatGPT 5.5 and Claude Opus 4.7 for language
improvement and editing. The data for this project was collected in the context of the \emph{bidt digitalbarometer.international
2026} funded by the Bavarian Research Institute for Digital Transformation (bidt).
Adrian Rauchleisch's work was supported by the National Science and Technology Council, Taiwan (R.O.C.) (grant no. 114-2628-H-002-007) and by the Taiwan Social Resilience Research Center (grant no. 115L9003) from the Higher Education Sprout Project by the Ministry of Education in Taiwan.

\printbibliography[
  heading=bibintoc,
  title={References}
]

\end{refsection}


\clearpage
\appendix

\renewcommand{\partname}{Supplementary Information}
\renewcommand{\thepart}{}

\part{}
\parttoc

\begin{refsection}

\section{Data}
\label{app:data}

\subsection{Power analysis}
Before starting the study, we performed power calculations using the \texttt{pwr} package in \texttt{R}. These calculations indicate that, with \(n = 1{,}000\) per country, one focal predictor, and eight covariates, the study has 90\% power at \(\alpha = .05\) to detect an incremental effect size of \(f^2 \approx 0.01\) in a standard individual-level regression framework. These calculations provide a useful benchmark for the individual-level associations that are central to our main analyses. They are less directly informative for indicators of cross-national heterogeneity, such as country-level variance components or random slopes, which depend more strongly on the number of countries than on the within-country sample size.

\subsection{Sample and data-quality checks}
\label{section:pop}
Data were collected using computer-assisted web interviews (CAWI) administered by Ipsos and its subsidiary Ipsos Interactive Services via the Ipsos Online Access Panel. Fieldwork was conducted between January 28, 2026, and March 2, 2026.

The target population comprised the internet-using population aged 18 years and older in each country.

Online access panels consist of pre-recruited individuals for whom key sociodemographic characteristics are known in advance, enabling targeted survey invitations. Upon successful completion of the survey, respondents received compensation in the form of points, allocated based on survey length and complexity.

To ensure sample quality, a quota-based sampling design was implemented to approximate the distribution of the internet-using adult population in each country. Quotas were applied for gender, age, and region (NUTS1), supplemented by a monitoring quota for education. Target distributions were based on official statistics (from Eurostat, United States Census Bureau, National Statistics of the Republic of China (Taiwan) and the United Nations Statistics Division). Given structural limitations of online panels in capturing very old age groups, age quotas were enforced up to 75 years, complemented by an open quota for older respondents.

Despite inherent limitations of online panels, particularly the undercoverage of less internet-affine populations, the realized samples closely matched the target sociodemographic distributions across countries. Minor deviations occurred in China and Taiwan, where quotas were slightly adjusted to ensure timely completion of fieldwork.

\subsection{Data Quality Assurance}

Data quality was ensured through multiple layers of validation and monitoring. During recruitment, initial verification procedures were conducted to validate respondent identities and exclude individuals exhibiting deceptive or inattentive response behavior. Additional safeguards included double opt-in procedures, geographic validation, detection of anonymous proxy servers, CAPTCHA verification, and duplicate response checks.

During fieldwork, respondent behavior was continuously monitored in real time using automated, self-adjusting algorithms designed to detect and exclude low-quality responses. These procedures included the identification of ``speeders'', defined as respondents completing the survey at less than one-third of the median completion time (following ESOMAR standards), and ``straightliners'', defined as respondents exhibiting invariant response patterns across grid questions, either in combination with unusually fast completion times or when grids contained reverse-coded items designed to elicit differentiated responses. Detection of such response patterns was fully automated and dynamically calibrated without reliance on fixed benchmarks. In total, 1,388 respondents were excluded based on these criteria.

In addition, three attention checks were embedded throughout the survey instrument: one open-ended item, one concealed within a grid question, and one simple single-choice item. Respondents who failed two of the three checks were excluded from the final dataset. This resulted in the removal of an additional 241 respondents across all countries.

\subsection{Pretesting and Field Procedures}

A pretest was conducted between January 28 and February 2, 2026, to verify questionnaire programming, filter logic, and overall survey functionality. The pretest confirmed the correct implementation of the instrument, and no modifications were required prior to full field deployment.

All stages of data collection and processing adhered to established quality standards. Ipsos maintains strict panel management guidelines and continuous monitoring procedures following participant recruitment to ensure sustained data quality. To minimize panel conditioning and respondent fatigue, panelists received no more than three to five survey invitations per month on average.

\section{Measures}
In this section, we report all item-level descriptive statistics and the wording of all questions used across all nine countries.
\subsection{Questionnaire Translation}

The questionnaires were originally developed in German and translated by Ipsos' in-house translation team, which specializes in survey translation. The source instrument was first translated into English, which then served as the basis for subsequent translations into the respective target languages.

These included country-specific adaptations, such as British English for the United Kingdom and American English for the United States, as well as Simplified Chinese for China and Traditional Chinese for Taiwan.

Following implementation of the translations in the programmed questionnaire, an independent review was conducted by linguists from the Ipsos team who had not been involved in the initial translation process, in order to ensure the accuracy and contextual appropriateness of the final survey instruments.

\subsection{Descriptive statistics: Germany}

\begin{longtable}{p{4.3cm}p{7.2cm}cc}
\caption{Descriptive statistics: Germany.}
\label{tab:descr_de} \\
\toprule
Variable & Question/Operationalization & M (SD) & n\\
\midrule
\endfirsthead

\multicolumn{4}{l}{\textit{Table \ref{tab:descr_de} continued from previous page}} \\
\toprule
Variable & Question/Operationalization & M (SD) & n\\
\midrule
\endhead

\midrule
\multicolumn{4}{r}{\textit{Continued on next page}} \\
\endfoot

\bottomrule
\endlastfoot

Support for military AI (6 items) 
& (1 = ``\"uberhaupt nicht angemessen'', 7 = ``v\"ollig angemessen'') 
& 4.83 (1.42) & 668\\
 & Das Militär setzt KI ein, um Satellitenbilder auszuwerten und feindliche Nachschubrouten zu erkennen. 
 & 5.33 (1.61) & 795\\
 & KI-Systeme analysieren aktuelle Gefechtsinformationen und empfehlen sichere Evakuierungsrouten. 
 & 5.10 (1.62) & 768\\
 & Ein KI-System steuert Drohnen, die feindliche Kommunikationsverbindungen gezielt stören. 
 & 4.91 (1.76) & 764\\
 & KI wird eingesetzt, um mögliche Ziele für künftige Luftangriffe zu identifizieren. 
 & 4.92 (1.76) & 765\\
 & KI-Systeme unterbreiten während laufender Gefechte Vorschläge für Angriffsoptionen auf Basis von Echtzeitdaten. 
 & 4.80 (1.66) & 751\\
 & Ein KI-gesteuertes Verteidigungssystem erkennt und zerstört automatisch ohne menschliches Eingreifen feindliche Einheiten auf dem Gefechtsfeld. 
 & 4.15 (1.91) & 754\\

\addlinespace[6pt]

Principled opposition to lethal autonomy (3 items) 
& (1 = ``stimme ganz und gar nicht zu'', 7 = ``stimme voll und ganz zu'') 
& 5.79 (1.36) & 818\\
 & Maschinen sollten niemals ohne menschliche Kontrolle über Leben und Tod entscheiden dürfen. 
 & 6.17 (1.41) & 934\\
 & Autonome Waffen verstoßen gegen grundlegende ethische Normen. 
 & 5.58 (1.69) & 842\\
 & Der Einsatz tödlicher autonomer Waffen-Systeme ist moralisch nicht vertretbar. 
 & 5.65 (1.72) & 889\\

\addlinespace[6pt]

AI benefits (4 items) 
& (1 = ``stimme ganz und gar nicht zu'', 7 = ``stimme voll und ganz zu'') 
& 4.25 (1.38) & 733\\
 & KI wird wesentlich zum deutschen Wirtschaftswachstum beitragen. 
 & 4.44 (1.61) & 858\\
 & KI wird der deutschen Regierung dabei helfen, effizienter für die Zukunft zu planen und Krisen zu bewältigen. 
 & 4.06 (1.71) & 852\\
 & KI hilft Parteien, erfolgreicher mit Wählern zu kommunizieren. 
 & 4.14 (1.75) & 825\\
 & KI wird der Menschheit helfen, existenzielle Bedrohungen erfolgreich zu bewältigen. 
 & 4.15 (1.67) & 843\\

\addlinespace[6pt]

AI risks (4 items) 
& (1 = ``stimme ganz und gar nicht zu'', 7 = ``stimme voll und ganz zu'') 
& 4.91 (1.22) & 798\\
 & KI wird viele Arbeitsplätze vernichten und zu großer Arbeitslosigkeit führen. 
 & 4.49 (1.70) & 906\\
 & Wir riskieren, die Kontrolle über unser Leben zu verlieren, wenn KI immer mehr Entscheidungen übernimmt. 
 & 5.05 (1.65) & 927\\
 & KI wird es den Parteien ermöglichen, Wähler erfolgreicher zu manipulieren. 
 & 5.11 (1.63) & 865\\
 & Unkontrollierte KI-Entwicklung wird eine existenzielle Bedrohung für die Menschheit darstellen. 
 & 5.03 (1.63) & 903\\

\addlinespace[6pt]

Hawk--dove orientations (4 items) 
& (1 = ``stimme ganz und gar nicht zu'', 7 = ``stimme voll und ganz zu''; last item reverse-coded) 
& 4.02 (1.34) & 783\\
 & Krieg ist manchmal notwendig, um die Interessen des eigenen Landes zu schützen. 
 & 3.43 (2.06) & 904\\
 & Deutschland braucht ein starkes Militär, um in der internationalen Politik etwas erreichen zu können. 
 & 4.90 (1.78) & 894\\
 & Deutschland sollte alles unternehmen, bis hin zum Einsatz militärischer Gewalt, um Angriffe expansionistischer Staaten zu verhindern. 
 & 4.60 (1.91) & 843\\
 & Der Einsatz militärischer Gewalt ist niemals gerechtfertigt. 
 & 5.03 (1.94) & 909\\

\addlinespace[6pt]

Trust in major powers (3 items) 
& (1 = ``\"uberhaupt nicht'', 7 = ``vollst\"andig'') 
& 2.42 (1.41) & 912\\
 & USA 
 & 2.61 (1.73) & 938\\
 & China 
 & 2.59 (1.65) & 919\\
 & Russland 
 & 2.11 (1.72) & 935\\

\addlinespace[6pt]

Confidence in international order (3 items) 
& (1 = ``stimme ganz und gar nicht zu'', 7 = ``stimme voll und ganz zu'') 
& 4.91 (1.31) & 845\\
 & Internationale Regeln und Zusammenarbeit tragen dazu bei, den Frieden zu sichern. 
 & 5.30 (1.56) & 931\\
 & Gemeinsame Regeln zwischen Staaten haben weltweit zu mehr Wohlstand beigetragen. 
 & 5.20 (1.49) & 886\\
 & Die internationale Ordnung ist in der Lage, globale Krisen wirksam zu bewältigen. 
 & 4.26 (1.74) & 882\\

\addlinespace[6pt]

Age 
& (in years) 
& 47.44 (15.89) & 1000\\
Gender 
& (1 = male) 
& 49.7\% & 1000\\
Education 
& (1 = Master's degree or higher) 
& 14.2\% & 1000\\

\end{longtable}

\subsection{Descriptive statistics: France}

\begin{longtable}{p{4.3cm}p{7.2cm}cc}
\caption{Descriptive statistics: France.}
\label{tab:descr_fr} \\
\toprule
Variable & Question/Operationalization & M (SD) & n\\
\midrule
\endfirsthead

\multicolumn{4}{l}{\textit{Table \ref{tab:descr_fr} continued from previous page}} \\
\toprule
Variable & Question/Operationalization & M (SD) & n\\
\midrule
\endhead

\midrule
\multicolumn{4}{r}{\textit{Continued on next page}} \\
\endfoot

\bottomrule
\endlastfoot

Support for military AI (6 items) 
& (1 = ``1 — Pas du tout approprié'', 7 = ``7 — Tout à fait approprié'') 
& 4.97 (1.31) & 567\\
 & Les forces armées utilisent l’IA pour analyser des images satellites et identifier les voies d’approvisionnement ennemies. 
 & 5.39 (1.42) & 734\\
 & Les systèmes d’IA analysent les informations de combat actuelles et recommandent des itinéraires d’évacuation sûrs. 
 & 5.08 (1.51) & 692\\
 & Un système d’IA contrôle des drones qui perturbent spécifiquement les communications ennemies. 
 & 5.06 (1.58) & 702\\
 & L’IA est utilisée pour identifier des cibles potentielles pour de futures frappes aériennes. 
 & 5.03 (1.60) & 706\\
 & Les systèmes d’IA font des suggestions d’options d’attaque basées sur des données en temps réel pendant les combats en cours. 
 & 4.93 (1.49) & 680\\
 & Un système de défense contrôlé par l’IA détecte et détruit automatiquement les unités ennemies sur le champ de bataille sans intervention humaine. 
 & 4.42 (1.86) & 674\\

\addlinespace[6pt]

Principled opposition to lethal autonomy (3 items) 
& (1 = ``1 — Complètement en désaccord'', 7 = ``7 — Complètement d’accord'') 
& 5.57 (1.38) & 687\\
 & Les machines ne devraient jamais être autorisées à prendre des décisions concernant la vie et la mort sans contrôle humain. 
 & 6.12 (1.46) & 904\\
 & Les armes autonomes violent les normes éthiques fondamentales. 
 & 5.31 (1.66) & 736\\
 & L’utilisation de systèmes d’armes autonomes létaux est moralement inacceptable. 
 & 5.39 (1.75) & 770\\

\addlinespace[6pt]

AI benefits (4 items) 
& (1 = ``1 — Complètement en désaccord'', 7 = ``7 — Complètement d’accord'') 
& 4.25 (1.47) & 637\\
 & L’IA stimulera une expansion économique significative en France. 
 & 4.29 (1.65) & 754\\
 & L’IA aidera les gouvernements à planifier l’avenir et à gérer les crises plus efficacement. 
 & 4.12 (1.80) & 778\\
 & L’IA aide les partis politiques à communiquer plus efficacement avec les électeurs. 
 & 4.24 (1.81) & 769\\
 & L’IA aidera l’humanité à faire face plus efficacement aux menaces existentielles. 
 & 4.12 (1.83) & 802\\

\addlinespace[6pt]

AI risks (4 items) 
& (1 = ``1 — Complètement en désaccord'', 7 = ``7 — Complètement d’accord'') 
& 5.43 (1.24) & 781\\
 & L’IA est susceptible d’entraîner des suppressions d’emplois et un chômage à grande échelle. 
 & 5.47 (1.47) & 902\\
 & À mesure que l’IA prend de plus en plus le relais dans la prise de décision, nous risquons de perdre le contrôle de nos vies. 
 & 5.31 (1.56) & 901\\
 & L’IA permettra aux partis politiques de manipuler plus efficacement les électeurs. 
 & 5.33 (1.63) & 838\\
 & Un développement incontrôlé de l’IA pourrait représenter des menaces existentielles pour l’humanité. 
 & 5.65 (1.50) & 898\\

\addlinespace[6pt]

Hawk--dove orientations (4 items) 
& (1 = ``1 — Complètement en désaccord'', 7 = ``7 — Complètement d’accord''; last item reverse-coded) 
& 4.80 (1.29) & 716\\
 & La guerre est parfois nécessaire pour protéger les intérêts de son pays. 
 & 4.53 (1.98) & 876\\
 & La France a besoin d’une armée forte pour réaliser quoi que ce soit en politique internationale. 
 & 5.51 (1.57) & 880\\
 & La France devrait faire tout ce qui est en son pouvoir, y compris recourir à la force militaire, pour empêcher les attaques des États expansionnistes. 
 & 5.03 (1.69) & 807\\
 & Le recours à la force militaire n’est jamais justifié. 
 & 4.08 (1.90) & 831\\

\addlinespace[6pt]

Trust in major powers (3 items) 
& (1 = ``1 — Pas du tout'', 7 = ``7 — Tout à fait'') 
& 2.43 (1.42) & 885\\
 & États-Unis 
 & 2.76 (1.79) & 917\\
 & Chine 
 & 2.51 (1.66) & 906\\
 & Russie 
 & 2.08 (1.68) & 920\\

\addlinespace[6pt]

Confidence in international order (3 items) 
& (1 = ``1 — Complètement en désaccord'', 7 = ``7 — Complètement d’accord'') 
& 4.32 (1.48) & 774\\
 & Les règles et la coopération internationales contribuent à garantir la paix. 
 & 4.68 (1.70) & 859\\
 & Les règles communes entre les pays ont contribué à une plus grande prospérité dans le monde entier. 
 & 4.54 (1.68) & 833\\
 & L’ordre international est capable de gérer efficacement les crises mondiales. 
 & 3.76 (1.71) & 828\\

\addlinespace[6pt]

Age 
& (in years) 
& 47.30 (16.30) & 1000\\
Gender 
& (1 = male) 
& 48.9\% & 1000\\
Education 
& (1 = Master's degree or higher) 
& 11.8\% & 1000\\

\end{longtable}

\subsection{Descriptive statistics: UK}

\begin{longtable}{p{4.3cm}p{7.2cm}cc}
\caption{Descriptive statistics: UK.}
\label{tab:descr_uk} \\
\toprule
Variable & Question/Operationalization & M (SD) & n\\
\midrule
\endfirsthead

\multicolumn{4}{l}{\textit{Table \ref{tab:descr_uk} continued from previous page}} \\
\toprule
Variable & Question/Operationalization & M (SD) & n\\
\midrule
\endhead

\midrule
\multicolumn{4}{r}{\textit{Continued on next page}} \\
\endfoot

\bottomrule
\endlastfoot

Support for military AI (6 items) 
& (1 = ``Not at all appropriate'', 7 = ``Completely appropriate'') 
& 4.61 (1.46) & 666\\
 & The military uses AI to analyse satellite images and identify enemy supply routes. 
 & 5.08 (1.61) & 782\\
 & AI systems analyse current combat information and recommend safe evacuation routes. 
 & 4.94 (1.62) & 779\\
 & An AI system controls drones that specifically disrupt enemy communications. 
 & 4.71 (1.68) & 788\\
 & AI is used to identify potential targets for future air strikes. 
 & 4.67 (1.79) & 787\\
 & AI systems make suggestions for attack options based on real-time data during ongoing combat. 
 & 4.45 (1.71) & 756\\
 & An AI-controlled defence system automatically detects and destroys enemy units on the battlefield without human intervention. 
 & 3.89 (1.94) & 767\\

\addlinespace[6pt]

Principled opposition to lethal autonomy (3 items) 
& (1 = ``Strongly disagree'', 7 = ``Strongly agree'') 
& 5.70 (1.26) & 770\\
 & Machines should never be allowed to make decisions about life and death without human control. 
 & 6.13 (1.37) & 934\\
 & Autonomous weapons violate fundamental ethical norms. 
 & 5.51 (1.50) & 793\\
 & Using lethal autonomous weapon systems is morally unacceptable. 
 & 5.52 (1.63) & 847\\

\addlinespace[6pt]

AI benefits (4 items) 
& (1 = ``Strongly disagree'', 7 = ``Strongly agree'') 
& 4.37 (1.43) & 677\\
 & AI will drive significant economic expansion in the United Kingdom. 
 & 4.51 (1.58) & 796\\
 & AI will help governments to more efficiently plan for the future and manage crises. 
 & 4.42 (1.60) & 828\\
 & AI helps parties to communicate more successfully with voters. 
 & 4.07 (1.72) & 789\\
 & AI will help humanity to address existential threats more successfully. 
 & 4.24 (1.68) & 804\\

\addlinespace[6pt]

AI risks (4 items) 
& (1 = ``Strongly disagree'', 7 = ``Strongly agree'') 
& 5.39 (1.12) & 783\\
 & AI is likely to cause widespread job displacement and unemployment. 
 & 5.36 (1.40) & 911\\
 & As AI increasingly takes over decision-making, we risk losing control over our lives. 
 & 5.34 (1.46) & 915\\
 & AI will allow parties to manipulate voters more successfully. 
 & 5.28 (1.46) & 843\\
 & Unchecked AI development could pose existential threats to humanity. 
 & 5.60 (1.42) & 902\\

\addlinespace[6pt]

Hawk--dove orientations (4 items) 
& (1 = ``Strongly disagree'', 7 = ``Strongly agree''; last item reverse-coded) 
& 4.94 (1.23) & 782\\
 & War is sometimes necessary to protect the interests of one's own country. 
 & 4.77 (1.79) & 880\\
 & The United Kingdom needs a strong military in order to achieve anything in international politics. 
 & 5.27 (1.54) & 887\\
 & The United Kingdom should do everything in its power, including the use of military force, to prevent attacks by expansionist states. 
 & 5.35 (1.43) & 863\\
 & The use of military force is never justified. 
 & 3.66 (1.87) & 878\\

\addlinespace[6pt]

Trust in major powers (3 items) 
& (1 = ``Not at all'', 7 = ``Completely'') 
& 2.52 (1.35) & 916\\
 & USA 
 & 2.92 (1.82) & 942\\
 & China 
 & 2.67 (1.64) & 920\\
 & Russia 
 & 1.95 (1.54) & 944\\

\addlinespace[6pt]

Confidence in international order (3 items) 
& (1 = ``Strongly disagree'', 7 = ``Strongly agree'') 
& 4.87 (1.31) & 790\\
 & International rules and cooperation help to secure peace. 
 & 5.20 (1.52) & 916\\
 & Common rules between countries have contributed to greater prosperity worldwide. 
 & 5.05 (1.46) & 860\\
 & The international order is capable of effectively managing global crises. 
 & 4.31 (1.64) & 823\\

\addlinespace[6pt]

Age 
& (in years) 
& 45.61 (16.16) & 1000\\
Gender 
& (1 = male) 
& 49.4\% & 1000\\
Education 
& (1 = Master's degree or higher) 
& 14.9\% & 1000\\

\end{longtable}

\subsection{Descriptive statistics: Italy}

\begin{longtable}{p{4.3cm}p{7.2cm}cc}
\caption{Descriptive statistics: Italy.}
\label{tab:descr_it} \\
\toprule
Variable & Question/Operationalization & M (SD) & n\\
\midrule
\endfirsthead

\multicolumn{4}{l}{\textit{Table \ref{tab:descr_it} continued from previous page}} \\
\toprule
Variable & Question/Operationalization & M (SD) & n\\
\midrule
\endhead

\midrule
\multicolumn{4}{r}{\textit{Continued on next page}} \\
\endfoot

\bottomrule
\endlastfoot

Support for military AI (6 items) 
& (1 = ``Per niente appropriato'', 7 = ``Completamente appropriato'') 
& 5.00 (1.31) & 647\\
 & L'IA viene utilizzata dalle forze armate per analizzare immagini satellitari e identificare le rotte di approvvigionamento del nemico. 
 & 5.19 (1.48) & 788\\
 & I sistemi basati sull'IA analizzano le informazioni di combattimento e raccomandano percorsi di evacuazione sicuri. 
 & 5.10 (1.47) & 783\\
 & Un sistema basato sull'IA controlla droni col compito specifico di interrompere le comunicazioni del nemico. 
 & 5.14 (1.46) & 773\\
 & L'IA viene utilizzata per identificare target potenziali per futuri attacchi aerei. 
 & 5.04 (1.61) & 756\\
 & I sistemi basati sull'IA suggeriscono obiettivi da attaccare in base a dati in tempo reale durante scontri in corso. 
 & 4.93 (1.56) & 744\\
 & Un sistema di difesa controllato dall'IA rileva e distrugge automaticamente unità nemiche sul campo di battaglia senza intervento umano. 
 & 4.65 (1.67) & 747\\

\addlinespace[6pt]

Principled opposition to lethal autonomy (3 items) 
& (1 = ``Completamente in disaccordo'', 7 = ``Completamente d’accordo'') 
& 5.82 (1.26) & 837\\
 & Alle macchine non dovrebbe mai essere consentito di prendere decisioni su vita o morte senza il controllo umano. 
 & 6.14 (1.29) & 940\\
 & Le armi autonome violano principi etici fondamentali. 
 & 5.64 (1.53) & 865\\
 & Utilizzare sistemi d'arma autonomi letali è moralmente inaccettabile. 
 & 5.73 (1.54) & 893\\

\addlinespace[6pt]

AI benefits (4 items) 
& (1 = ``Completamente in disaccordo'', 7 = ``Completamente d’accordo'') 
& 4.45 (1.43) & 714\\
 & L'IA genererà una notevole espansione economica in Italia. 
 & 4.45 (1.63) & 836\\
 & L'IA aiuterà i governi a pianificare per il futuro e a gestire le crisi in modo più efficiente. 
 & 4.43 (1.71) & 831\\
 & L'IA aiuta i partiti a comunicare meglio con gli elettori. 
 & 4.15 (1.82) & 820\\
 & L'IA aiuterà l'umanità ad affrontare con più successo minacce alla sopravvivenza. 
 & 4.50 (1.65) & 838\\

\addlinespace[6pt]

AI risks (4 items) 
& (1 = ``Completamente in disaccordo'', 7 = ``Completamente d’accordo'') 
& 4.99 (1.34) & 771\\
 & L'IA probabilmente causerà perdite di posti di lavoro e disoccupazione su vasta scala. 
 & 5.10 (1.55) & 897\\
 & Con l'aumento dell'uso dell'IA per prendere decisioni, rischiamo di perdere il controllo delle nostre vite. 
 & 5.12 (1.56) & 904\\
 & L'IA consentirà ai partiti di manipolare gli elettori con maggiore successo. 
 & 4.91 (1.74) & 848\\
 & Lo sviluppo incontrollato dell'IA potrebbe porre minacce alla sopravvivenza umana. 
 & 4.85 (1.72) & 872\\

\addlinespace[6pt]

Hawk--dove orientations (4 items) 
& (1 = ``Completamente in disaccordo'', 7 = ``Completamente d’accordo''; last item reverse-coded) 
& 3.72 (1.43) & 800\\
 & A volte la guerra è necessaria per proteggere gli interessi del proprio Paese. 
 & 3.60 (2.07) & 902\\
 & L'Italia ha bisogno di forze armate potenti per ottenere qualsiasi cosa nella politica internazionale. 
 & 3.87 (2.01) & 869\\
 & L'Italia dovrebbe fare tutto ciò che può, compreso utilizzare forza militare, per prevenire attacchi da parte di Paesi espansionistici. 
 & 4.50 (1.90) & 869\\
 & L'uso della forza militare non è mai giustificato. 
 & 5.22 (1.84) & 917\\

\addlinespace[6pt]

Trust in major powers (3 items) 
& (1 = ``Per nulla'', 7 = ``Completamente'') 
& 2.91 (1.57) & 898\\
 & Stati Uniti 
 & 3.01 (1.88) & 925\\
 & Cina 
 & 3.14 (1.82) & 914\\
 & Russia 
 & 2.62 (1.94) & 922\\

\addlinespace[6pt]

Confidence in international order (3 items) 
& (1 = ``Completamente in disaccordo'', 7 = ``Completamente d’accordo'') 
& 4.79 (1.33) & 857\\
 & Le regole e la cooperazione internazionale aiutano ad assicurare la pace. 
 & 5.15 (1.58) & 925\\
 & Regole comuni tra Paesi hanno contribuito a una maggiore prosperità in tutto il mondo. 
 & 5.08 (1.49) & 904\\
 & L'ordine internazionale è in grado di gestire efficacemente le crisi globali. 
 & 4.09 (1.74) & 881\\

\addlinespace[6pt]

Age 
& (in years) 
& 48.43 (15.66) & 1000\\
Gender 
& (1 = male) 
& 49.7\% & 1000\\
Education 
& (1 = Master's degree or higher) 
& 15.3\% & 1000\\

\end{longtable}

\subsection{Descriptive statistics: Spain}

\begin{longtable}{p{4.3cm}p{7.2cm}cc}
\caption{Descriptive statistics: Spain.}
\label{tab:descr_es} \\
\toprule
Variable & Question/Operationalization & M (SD) & n\\
\midrule
\endfirsthead

\multicolumn{4}{l}{\textit{Table \ref{tab:descr_es} continued from previous page}} \\
\toprule
Variable & Question/Operationalization & M (SD) & n\\
\midrule
\endhead

\midrule
\multicolumn{4}{r}{\textit{Continued on next page}} \\
\endfoot

\bottomrule
\endlastfoot

Support for military AI (6 items) 
& (1 = ``Nada adecuado'', 7 = ``Completamente adecuado'') 
& 5.15 (1.21) & 694\\
 & El ejército utiliza la IA para analizar imágenes por satélite e identificar las rutas de abastecimiento del enemigo. 
 & 5.30 (1.42) & 822\\
 & Los sistemas de IA analizan la información actual sobre el combate y recomiendan rutas de evacuación seguras. 
 & 5.20 (1.41) & 813\\
 & Un sistema de inteligencia artificial controla drones que interfieren específicamente las comunicaciones enemigas. 
 & 5.21 (1.45) & 825\\
 & La IA se utiliza para identificar posibles objetivos para futuros ataques aéreos. 
 & 5.22 (1.45) & 814\\
 & Los sistemas de IA sugieren opciones de ataque basadas en datos en tiempo real durante el combate. 
 & 5.12 (1.48) & 793\\
 & Un sistema de defensa controlado por IA detecta y destruye automáticamente las unidades enemigas en el campo de batalla sin intervención humana. 
 & 4.87 (1.65) & 796\\

\addlinespace[6pt]

Principled opposition to lethal autonomy (3 items) 
& (1 = ``Totalmente en desacuerdo'', 7 = ``Totalmente de acuerdo'') 
& 5.77 (1.20) & 832\\
 & Nunca se debe permitir que las máquinas tomen decisiones sobre la vida y la muerte sin control humano. 
 & 6.09 (1.31) & 961\\
 & Las armas autónomas violan las normas éticas fundamentales. 
 & 5.58 (1.48) & 859\\
 & El uso de sistemas de armas autónomas letales es moralmente inaceptable. 
 & 5.71 (1.49) & 900\\

\addlinespace[6pt]

AI benefits (4 items) 
& (1 = ``Totalmente en desacuerdo'', 7 = ``Totalmente de acuerdo'') 
& 4.69 (1.38) & 773\\
 & La IA impulsará una importante expansión económica en España. 
 & 4.76 (1.58) & 872\\
 & La IA ayudará a los gobiernos a planificar el futuro y gestionar las crisis de manera más eficiente. 
 & 4.67 (1.59) & 883\\
 & La IA ayuda a los partidos a comunicarse mejor con los votantes. 
 & 4.46 (1.78) & 868\\
 & La IA ayudará a la humanidad a abordar las amenazas existenciales con mayor éxito. 
 & 4.73 (1.63) & 875\\

\addlinespace[6pt]

AI risks (4 items) 
& (1 = ``Totalmente en desacuerdo'', 7 = ``Totalmente de acuerdo'') 
& 5.28 (1.15) & 825\\
 & Es probable que la IA provoque un desplazamiento generalizado de puestos de trabajo y desempleo. 
 & 5.34 (1.41) & 944\\
 & A medida que la IA adquiere cada vez más protagonismo en la toma de decisiones, corremos el riesgo de perder el control sobre nuestras vidas. 
 & 5.19 (1.49) & 945\\
 & La IA permitirá a los partidos manipular a los votantes con mayor éxito. 
 & 5.24 (1.60) & 869\\
 & El desarrollo descontrolado de la IA podría suponer una amenaza existencial para la humanidad. 
 & 5.40 (1.50) & 944\\

\addlinespace[6pt]

Hawk--dove orientations (4 items) 
& (1 = ``Totalmente en desacuerdo'', 7 = ``Totalmente de acuerdo''; last item reverse-coded) 
& 4.27 (1.31) & 803\\
 & A veces la guerra es necesaria para proteger los intereses del propio país. 
 & 4.15 (2.08) & 910\\
 & España necesita un ejército fuerte para lograr algo en la política internacional. 
 & 4.92 (1.80) & 909\\
 & España debería hacer todo lo que esté en su mano, incluido el uso de la fuerza militar, para impedir los ataques de los Estados expansionistas. 
 & 4.88 (1.74) & 880\\
 & El uso de la fuerza militar nunca está justificado. 
 & 4.93 (1.88) & 921\\

\addlinespace[6pt]

Trust in major powers (3 items) 
& (1 = ``Nada en absoluto'', 7 = ``Completamente'') 
& 3.06 (1.59) & 900\\
 & Estados Unidos 
 & 3.07 (2.00) & 934\\
 & China 
 & 3.59 (1.89) & 917\\
 & Rusia 
 & 2.59 (1.86) & 933\\

\addlinespace[6pt]

Confidence in international order (3 items) 
& (1 = ``Totalmente en desacuerdo'', 7 = ``Totalmente de acuerdo'') 
& 4.87 (1.35) & 887\\
 & Las normas internacionales y la cooperación contribuyen a garantizar la paz. 
 & 5.07 (1.60) & 950\\
 & La existencia de normas comunes entre los países ha contribuido a una mayor prosperidad en todo el mundo. 
 & 5.11 (1.52) & 933\\
 & El orden internacional es capaz de gestionar eficazmente las crisis mundiales. 
 & 4.40 (1.69) & 910\\

\addlinespace[6pt]

Age 
& (in years) 
& 46.74 (15.37) & 1000\\
Gender 
& (1 = male) 
& 49.6\% & 1000\\
Education 
& (1 = Master's degree or higher) 
& 31.9\% & 1000\\

\end{longtable}

\subsection{Descriptive statistics: Finland}

\begin{longtable}{p{4.3cm}p{7.2cm}cc}
\caption{Descriptive statistics: Finland.}
\label{tab:descr_fi} \\
\toprule
Variable & Question/Operationalization & M (SD) & n\\
\midrule
\endfirsthead

\multicolumn{4}{l}{\textit{Table \ref{tab:descr_fi} continued from previous page}} \\
\toprule
Variable & Question/Operationalization & M (SD) & n\\
\midrule
\endhead

\midrule
\multicolumn{4}{r}{\textit{Continued on next page}} \\
\endfoot

\bottomrule
\endlastfoot

Support for military AI (6 items) 
& (1 = ``Ei lainkaan asianmukaista'', 7 = ``Täysin asianmukaista'') 
& 4.79 (1.36) & 756\\
 & Tekoälyn sotilaallinen käyttö satelliittikuvien analysointiin ja vihollisten toimitusreittien tunnistamiseen. 
 & 5.04 (1.55) & 878\\
 & Tekoälyjärjestelmillä analysoidaan ajankohtaista taistelutietoa ja suositellaan turvallisia evakuointireittejä. 
 & 5.02 (1.55) & 867\\
 & Tekoälyjärjestelmällä ohjataan lennokkeja, joilla häiritään erityisesti vihollisten viestintää. 
 & 4.98 (1.62) & 867\\
 & Tekoälyä käytetään mahdollisten tulevien ilmaiskukohteiden tunnistamiseen. 
 & 4.93 (1.57) & 855\\
 & Tekoälyjärjestelmät ehdottavat hyökkäysvaihtoehtoja meneillään olevan taistelun aikana reaaliaikaisen datan perusteella. 
 & 4.58 (1.65) & 838\\
 & Tekoälyn ohjaama puolustusjärjestelmä havaitsee ja tuhoaa automaattisesti taistelukentällä olevat vihollisyksiköt ilman ihmisten väliintuloa. 
 & 4.22 (1.78) & 846\\

\addlinespace[6pt]

Principled opposition to lethal autonomy (3 items) 
& (1 = ``Täysin eri mieltä'', 7 = ``Täysin samaa mieltä'') 
& 5.57 (1.29) & 827\\
 & Koneiden ei pitäisi koskaan antaa tehdä päätöksiä elämästä ja kuolemasta ilman ihmisten valvontaa. 
 & 6.12 (1.33) & 958\\
 & Autonomiset aseet rikkovat perustavanlaatuisia eettisiä perusnormeja. 
 & 5.25 (1.59) & 849\\
 & Tappavien autonomisten asejärjestelmien käyttö ei ole moraalisesti hyväksyttävää. 
 & 5.43 (1.58) & 885\\

\addlinespace[6pt]

AI benefits (4 items) 
& (1 = ``Täysin eri mieltä'', 7 = ``Täysin samaa mieltä'') 
& 4.11 (1.36) & 741\\
 & Tekoäly edistää merkittävää talouskasvua maassa Suomessa. 
 & 4.22 (1.57) & 874\\
 & Tekoäly auttaa hallituksia suunnittelemaan tehokkaammin tulevaisuutta ja hallitsemaan kriisejä. 
 & 4.08 (1.58) & 875\\
 & Tekoäly auttaa puolueita kommunikoimaan menestyksekkäämmin äänestäjien kanssa. 
 & 3.93 (1.62) & 853\\
 & Tekoäly auttaa ihmiskuntaa vastaamaan eksistentiaalisiin uhkiin menestyksekkäämmin. 
 & 4.13 (1.58) & 838\\

\addlinespace[6pt]

AI risks (4 items) 
& (1 = ``Täysin eri mieltä'', 7 = ``Täysin samaa mieltä'') 
& 5.07 (1.17) & 809\\
 & Tekoäly aiheuttaa todennäköisesti laajamittaisesti työpaikkojen menetyksiä ja työttömyyttä. 
 & 4.94 (1.51) & 923\\
 & Koska tekoäly ottaa enenevässä määrin vastuuta päätöksenteosta, olemme vaarassa menettää elämämme hallinnan. 
 & 4.74 (1.62) & 926\\
 & Tekoälyn avulla puolueet voivat manipuloida äänestäjiä menestyksekkäämmin. 
 & 5.08 (1.48) & 885\\
 & Valvomaton tekoälyn kehitys voi aiheuttaa eksistentiaalisia uhkia ihmiskunnalle. 
 & 5.54 (1.44) & 925\\

\addlinespace[6pt]

Hawk--dove orientations (4 items) 
& (1 = ``Täysin eri mieltä'', 7 = ``Täysin samaa mieltä''; last item reverse-coded) 
& 4.81 (1.20) & 820\\
 & Sota on toisinaan tarpeen oman maan etujen suojelemiseksi. 
 & 4.38 (1.93) & 890\\
 & Suomi tarvitsee vahvat puolustusvoimat, jotta se voi saavuttaa jotakin kansainvälisessä politiikassa. 
 & 5.36 (1.57) & 919\\
 & Suomen tulee tehdä voitavansa, mukaan lukien käyttää sotilaallista voimaa, estääkseen laajentumaan pyrkivien valtioiden hyökkäykset. 
 & 5.55 (1.44) & 905\\
 & Sotilaallisen voiman käyttö ei ole koskaan oikeutettua. 
 & 4.11 (2.01) & 910\\

\addlinespace[6pt]

Trust in major powers (3 items) 
& (1 = ``En lainkaan'', 7 = ``Täysin'') 
& 2.50 (1.25) & 936\\
 & Yhdysvallat 
 & 2.92 (1.66) & 951\\
 & Kiina 
 & 2.80 (1.53) & 945\\
 & Venäjä 
 & 1.79 (1.42) & 958\\

\addlinespace[6pt]

Confidence in international order (3 items) 
& (1 = ``Täysin eri mieltä'', 7 = ``Täysin samaa mieltä'') 
& 5.13 (1.25) & 895\\
 & Kansainväliset säännöt ja yhteistyö auttavat turvaamaan rauhan. 
 & 5.17 (1.47) & 945\\
 & Yhteiset säännöt maiden välillä ovat osaltaan lisänneet hyvinvointia maailmanlaajuisesti. 
 & 5.19 (1.39) & 921\\
 & Kansainvälisen järjestyksen avulla pystytään hallitsemaan tehokkaasti globaaleja kriisejä. 
 & 4.99 (1.42) & 919\\

\addlinespace[6pt]

Age 
& (in years) 
& 46.64 (16.57) & 1000\\
Gender 
& (1 = male) 
& 49.8\% & 1000\\
Education 
& (1 = Master's degree or higher) 
& 15.2\% & 1000\\

\end{longtable}

\subsection{Descriptive statistics: USA}

\begin{longtable}{p{4.3cm}p{7.2cm}cc}
\caption{Descriptive statistics: USA.}
\label{tab:descr_us} \\
\toprule
Variable & Question/Operationalization & M (SD) & n\\
\midrule
\endfirsthead

\multicolumn{4}{l}{\textit{Table \ref{tab:descr_us} continued from previous page}} \\
\toprule
Variable & Question/Operationalization & M (SD) & n\\
\midrule
\endhead

\midrule
\multicolumn{4}{r}{\textit{Continued on next page}} \\
\endfoot

\bottomrule
\endlastfoot

Support for military AI (6 items) 
& (1 = ``Not at all appropriate'', 7 = ``Completely appropriate'') 
& 4.88 (1.43) & 689\\
 & The military uses AI to analyze satellite images and identify enemy supply routes. 
 & 5.22 (1.58) & 802\\
 & AI systems analyze current combat information and recommend safe evacuation routes. 
 & 5.15 (1.59) & 803\\
 & An AI system controls drones that specifically disrupt enemy communications. 
 & 4.99 (1.71) & 792\\
 & AI is used to identify potential targets for future air strikes. 
 & 4.85 (1.77) & 789\\
 & AI systems make suggestions for attack options based on real-time data during ongoing combat. 
 & 4.83 (1.73) & 779\\
 & An AI-controlled defense system automatically detects and destroys enemy units on the battlefield without human intervention. 
 & 4.31 (1.97) & 779\\

\addlinespace[6pt]

Principled opposition to lethal autonomy (3 items) 
& (1 = ``Strongly disagree'', 7 = ``Strongly agree'') 
& 5.44 (1.37) & 777\\
 & Machines should never be allowed to make decisions about life and death without human control. 
 & 6.05 (1.39) & 928\\
 & Autonomous weapons violate fundamental ethical norms. 
 & 5.16 (1.72) & 813\\
 & Using lethal autonomous weapon systems is morally unacceptable. 
 & 5.15 (1.80) & 826\\

\addlinespace[6pt]

AI benefits (4 items) 
& (1 = ``Strongly disagree'', 7 = ``Strongly agree'') 
& 4.55 (1.49) & 698\\
 & AI will drive significant economic expansion in the United States. 
 & 4.68 (1.70) & 822\\
 & AI will help governments to more efficiently plan for the future and manage crises. 
 & 4.45 (1.77) & 830\\
 & AI helps parties to communicate more successfully with voters. 
 & 4.25 (1.88) & 795\\
 & AI will help humanity to address existential threats more successfully. 
 & 4.44 (1.72) & 820\\

\addlinespace[6pt]

AI risks (4 items) 
& (1 = ``Strongly disagree'', 7 = ``Strongly agree'') 
& 5.30 (1.30) & 769\\
 & AI is likely to cause widespread job displacement and unemployment. 
 & 5.23 (1.61) & 913\\
 & As AI increasingly takes over decision-making, we risk losing control over our lives. 
 & 5.27 (1.62) & 907\\
 & AI will allow parties to manipulate voters more successfully. 
 & 5.16 (1.67) & 822\\
 & Unchecked AI development could pose existential threats to humanity. 
 & 5.59 (1.53) & 894\\

\addlinespace[6pt]

Hawk--dove orientations (4 items) 
& (1 = ``Strongly disagree'', 7 = ``Strongly agree''; last item reverse-coded) 
& 4.89 (1.26) & 781\\
 & War is sometimes necessary to protect the interests of one's own country. 
 & 5.04 (1.74) & 900\\
 & The United States needs a strong military in order to achieve anything in international politics. 
 & 5.17 (1.69) & 893\\
 & The United States should do everything in its power, including the use of military force, to prevent attacks by expansionist states. 
 & 5.01 (1.68) & 836\\
 & The use of military force is never justified. 
 & 3.70 (2.01) & 881\\

\addlinespace[6pt]

Trust in major powers (3 items) 
& (1 = ``Not at all'', 7 = ``Completely'') 
& 3.35 (1.46) & 888\\
 & USA 
 & 4.64 (1.97) & 927\\
 & China 
 & 2.90 (1.84) & 907\\
 & Russia 
 & 2.51 (1.82) & 909\\

\addlinespace[6pt]

Confidence in international order (3 items) 
& (1 = ``Strongly disagree'', 7 = ``Strongly agree'') 
& 5.09 (1.27) & 797\\
 & International rules and cooperation help to secure peace. 
 & 5.38 (1.45) & 901\\
 & Common rules between countries have contributed to greater prosperity worldwide. 
 & 5.31 (1.40) & 868\\
 & The international order is capable of effectively managing global crises. 
 & 4.58 (1.69) & 825\\

\addlinespace[6pt]

Age 
& (in years) 
& 45.53 (16.43) & 1000\\
Gender 
& (1 = male) 
& 49.6\% & 1000\\
Education 
& (1 = Master's degree or higher) 
& 12.7\% & 1000\\

\end{longtable}

\subsection{Descriptive statistics: Taiwan}

\begin{longtable}{p{4.3cm}p{7.2cm}cc}
\caption{Descriptive statistics: Taiwan.}
\label{tab:descr_tw} \\
\toprule
Variable & Question/Operationalization & M (SD) & n\\
\midrule
\endfirsthead

\multicolumn{4}{l}{\textit{Table \ref{tab:descr_tw} continued from previous page}} \\
\toprule
Variable & Question/Operationalization & M (SD) & n\\
\midrule
\endhead

\midrule
\multicolumn{4}{r}{\textit{Continued on next page}} \\
\endfoot

\bottomrule
\endlastfoot

Support for military AI (6 items) 
& (1 = ``一點都不適合'', 7 = ``完全適合'') 
& 5.30 (1.02) & 857\\
 & 軍方使用 AI 分析衛星影像，並找出敵人補給路線。 
 & 5.40 (1.25) & 915\\
 & AI 系統分析當下的戰鬥資訊，並建議安全的疏散路線。 
 & 5.33 (1.23) & 909\\
 & AI 系統控制專門干擾敵人通訊的無人機。 
 & 5.36 (1.26) & 907\\
 & 使用 AI 找出未來空襲的潛在目標。 
 & 5.38 (1.25) & 904\\
 & AI 系統會根據持續戰鬥期間的即時資料，提供攻擊選項建議。 
 & 5.28 (1.25) & 908\\
 & AI 控制的防禦系統在沒有人為介入的情況下，自動偵測並摧毀戰場上的敵人單位。 
 & 5.06 (1.37) & 902\\

\addlinespace[6pt]

Principled opposition to lethal autonomy (3 items) 
& (1 = ``非常不同意'', 7 = ``非常同意'') 
& 5.37 (1.16) & 876\\
 & 若無人類控制，絕不可讓機器做出有關生殺大權的決策。 
 & 5.71 (1.36) & 950\\
 & 自主性武器違反基本道德規範。 
 & 5.10 (1.46) & 890\\
 & 使用致命性自主武器系統在道德上是不可接受的。 
 & 5.36 (1.45) & 915\\

\addlinespace[6pt]

AI benefits (4 items) 
& (1 = ``非常不同意'', 7 = ``非常同意'') 
& 5.00 (1.08) & 861\\
 & AI 將促使台灣經濟大幅成長。 
 & 5.23 (1.34) & 939\\
 & AI 將幫助政府更高效地規劃未來和進行危機管理。 
 & 5.05 (1.31) & 934\\
 & AI 能幫助政黨更有效地對選民溝通。 
 & 4.64 (1.52) & 893\\
 & AI 將幫助人類更有效地應對生存威脅。 
 & 5.03 (1.34) & 930\\

\addlinespace[6pt]

AI risks (4 items) 
& (1 = ``非常不同意'', 7 = ``非常同意'') 
& 5.12 (1.02) & 871\\
 & AI 可能會造成大量的工作被取代和失業問題。 
 & 5.29 (1.31) & 957\\
 & 隨著 AI 逐漸主導決策過程，我們可能面對失去對生活的控制的風險。 
 & 5.11 (1.30) & 943\\
 & AI 將使政黨更有能力操控選民。 
 & 4.81 (1.57) & 904\\
 & 未受控制的 AI 發展，可能對人類的生存構成威脅。 
 & 5.33 (1.36) & 947\\

\addlinespace[6pt]

Hawk--dove orientations (4 items) 
& (1 = ``非常不同意'', 7 = ``非常同意''; last item reverse-coded) 
& 4.30 (1.06) & 860\\
 & 為了保護自身國家的利益，有時候戰爭是必要的。 
 & 4.39 (1.91) & 928\\
 & 台灣需要強大的軍隊，才能在國際政治上成就任何功績。 
 & 4.90 (1.65) & 935\\
 & 台灣應盡其一切所能，包括運用軍事力量，以避免遭到擴張主義國家攻擊。 
 & 5.31 (1.55) & 936\\
 & 軍事力量的使用絕不能被合理化。 
 & 5.48 (1.38) & 926\\

\addlinespace[6pt]

Trust in major powers (3 items) 
& (1 = ``完全不信任'', 7 = ``完全信任'') 
& 3.19 (1.32) & 922\\
 & 美國 
 & 4.02 (1.76) & 953\\
 & 中國 
 & 2.85 (1.87) & 952\\
 & 俄羅斯 
 & 2.70 (1.72) & 933\\

\addlinespace[6pt]

Confidence in international order (3 items) 
& (1 = ``非常不同意'', 7 = ``非常同意'') 
& 5.14 (1.18) & 926\\
 & 國際規則與合作有助於確保和平。 
 & 5.25 (1.35) & 957\\
 & 國家/地區之間的共同規則對於增進全世界繁榮已有所貢獻。 
 & 5.16 (1.27) & 946\\
 & 國際秩序能夠有效管理全球危機。 
 & 4.99 (1.47) & 951\\

\addlinespace[6pt]

Age 
& (in years) 
& 43.30 (15.22) & 1000\\
Gender 
& (1 = male) 
& 49.0\% & 1000\\
Education 
& (1 = Master's degree or higher) 
& 14.1\% & 1000\\

\end{longtable}

\subsection{Descriptive statistics: China}

\begin{longtable}{p{4.3cm}p{7.2cm}cc}
\caption{Descriptive statistics: China.}
\label{tab:descr_cn} \\
\toprule
Variable & Question/Operationalization & M (SD) & n\\
\midrule
\endfirsthead

\multicolumn{4}{l}{\textit{Table \ref{tab:descr_cn} continued from previous page}} \\
\toprule
Variable & Question/Operationalization & M (SD) & n\\
\midrule
\endhead

\midrule
\multicolumn{4}{r}{\textit{Continued on next page}} \\
\endfoot

\bottomrule
\endlastfoot

Support for military AI (6 items) 
& (1 = ``完全不恰当'', 7 = ``极为恰当'') 
& 5.39 (0.90) & 827\\
 & 军方利用 AI 分析卫星图像并确定敌方补给路线。 
 & 5.50 (1.16) & 918\\
 & AI 系统分析当前战斗信息并推荐安全疏散路线。 
 & 5.47 (1.15) & 926\\
 & AI 系统控制无人机专门中断敌方通信。 
 & 5.52 (1.16) & 914\\
 & AI 用于确定未来空袭的潜在目标。 
 & 5.41 (1.25) & 902\\
 & AI 系统根据持续战斗期间的实时数据提出攻击方案建议。 
 & 5.40 (1.20) & 909\\
 & AI 控制的防御系统自动检测并摧毁战场上的敌方单位，而无需人为干预。 
 & 4.98 (1.43) & 903\\

\addlinespace[6pt]

Principled opposition to lethal autonomy (3 items) 
& (1 = ``强烈反对'', 7 = ``强烈赞同'') 
& 5.08 (1.29) & 880\\
 & 决不能让机器在没有人力控制的情况下判定他人生死。 
 & 5.49 (1.49) & 953\\
 & 自主武器违反基本道德规范。 
 & 4.80 (1.59) & 909\\
 & 使用致命的自主武器系统在道义上是不可接受的。 
 & 4.98 (1.58) & 914\\

\addlinespace[6pt]

AI benefits (4 items) 
& (1 = ``强烈反对'', 7 = ``强烈赞同'') 
& 5.24 (0.97) & 865\\
 & 人工智能将促使中国的经济大幅扩张。 
 & 5.36 (1.27) & 938\\
 & 人工智能将帮助政府更高效地规划未来和进行危机管理。 
 & 5.39 (1.26) & 940\\
 & 人工智能将帮助政党更有效地与选民进行沟通。 
 & 5.04 (1.39) & 905\\
 & 人工智能将帮助人类更有效地应对生存威胁。 
 & 5.14 (1.37) & 933\\

\addlinespace[6pt]

AI risks (4 items) 
& (1 = ``强烈反对'', 7 = ``强烈赞同'') 
& 4.56 (1.24) & 864\\
 & 人工智能可能会造成大量的工作被取代和失业问题。 
 & 4.67 (1.65) & 957\\
 & 随着人工智能逐渐主导决策过程，我们可能会失去对生活的掌控。 
 & 4.20 (1.72) & 938\\
 & 人工智能将使政党更有能力操控选民。 
 & 4.55 (1.67) & 897\\
 & 不受控制的人工智能发展可能会对人类的生存构成威胁。 
 & 4.89 (1.67) & 945\\

\addlinespace[6pt]

Hawk--dove orientations (4 items) 
& (1 = ``强烈反对'', 7 = ``强烈赞同''; last item reverse-coded) 
& 5.27 (0.98) & 908\\
 & 为了保护自己国家的利益，战争有时是必要的。 
 & 5.54 (1.37) & 949\\
 & 中国需要一支强大的军队才能在国际政治事务中取得任何成就。 
 & 5.72 (1.34) & 956\\
 & 中国应尽一切努力，包括使用军事力量，以防止扩张主义国家的攻击。 
 & 5.84 (1.26) & 958\\
 & 使用武力是没有道理的。 
 & 4.00 (1.84) & 941\\

\addlinespace[6pt]

Trust in major powers (3 items) 
& (1 = ``根本不相信'', 7 = ``完全相信'') 
& 4.63 (0.89) & 948\\
 & 美国 
 & 3.22 (1.91) & 965\\
 & 中国 
 & 6.17 (1.10) & 975\\
 & 俄罗斯 
 & 4.48 (1.47) & 951\\

\addlinespace[6pt]

Confidence in international order (3 items) 
& (1 = ``强烈反对'', 7 = ``强烈赞同'') 
& 5.66 (0.98) & 943\\
 & 国际规则与合作有助于确保和平。 
 & 5.81 (1.13) & 968\\
 & 各个国家/地区之间的共同规则促进了全世界的更大繁荣。 
 & 5.73 (1.15) & 963\\
 & 国际秩序能够有效地管理全球危机。 
 & 5.44 (1.29) & 951\\

\addlinespace[6pt]

Age 
& (in years) 
& 45.20 (15.43) & 1000\\
Gender 
& (1 = male) 
& 49.2\% & 1000\\
Education 
& (1 = Master's degree or higher) 
& 6.4\% & 1000\\

\end{longtable}

\subsection{Summary predictors}

\begin{figure}[H]
    \centering
    \includegraphics[width=\linewidth]{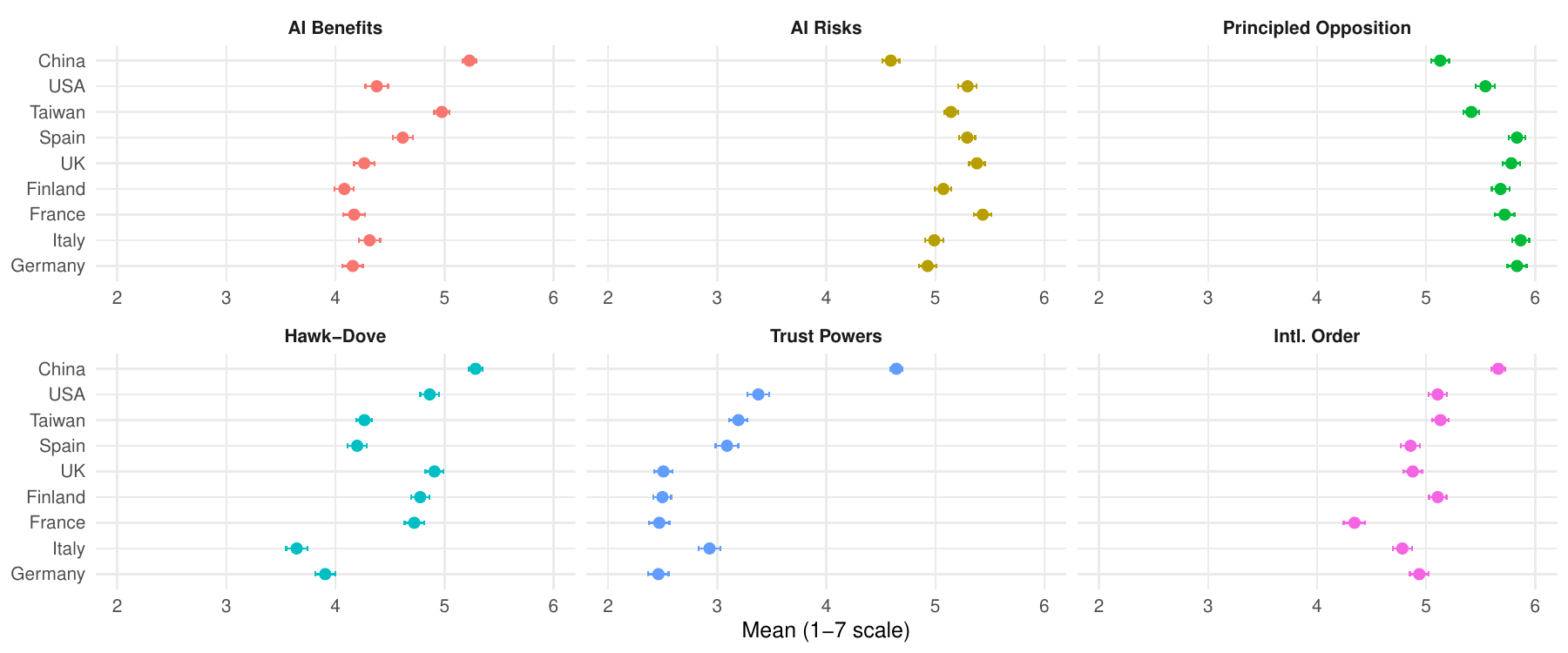}
    \caption{Country-level means and 95\% confidence intervals for all predictor indices. Means are computed on all available item
  responses prior to imputation; respondents contributed to an index if they answered at least one constituent item.}
    \label{fig:SI_Figure_01}
\end{figure}

\subsection{Scale reliabilities}

\begin{table}[H]
\centering
\resizebox{\textwidth}{!}{
\begin{tabular}{lccccccc}
\toprule
Country & Mil. AI support & Princ. opposition & AI benefits & AI risks & Hawk{--}dove & Trust powers & Intl. order\\
\midrule
Germany & 0.90 & 0.80 & 0.84 & 0.74 & 0.65 & 0.77 & 0.76\\
France & 0.90 & 0.80 & 0.85 & 0.82 & 0.70 & 0.78 & 0.84\\
UK & 0.91 & 0.78 & 0.88 & 0.78 & 0.72 & 0.73 & 0.81\\
Italy & 0.92 & 0.82 & 0.86 & 0.81 & 0.72 & 0.78 & 0.78\\
Spain & 0.90 & 0.79 & 0.86 & 0.78 & 0.66 & 0.78 & 0.80\\
Finland & 0.91 & 0.81 & 0.87 & 0.78 & 0.62 & 0.75 & 0.85\\
USA & 0.90 & 0.77 & 0.88 & 0.81 & 0.69 & 0.68 & 0.78\\
Taiwan & 0.90 & 0.74 & 0.80 & 0.71 & 0.58 & 0.58 & 0.83\\
China & 0.82 & 0.77 & 0.71 & 0.73 & 0.58 & 0.30 & 0.76\\
\midrule
\textbf{Overall} & 0.90 & 0.79 & 0.86 & 0.78 & 0.69 & 0.70 & 0.81\\
\bottomrule
\end{tabular}
}
\caption{Cronbach's $\alpha$ by country and scale.}
\label{tab:alpha_country}
\end{table}

\section{Outcome variable Military AI use}

\subsection{Validation of the military AI support scale}
We assessed the dimensional structure of the six military AI support items using confirmatory factor analysis (CFA) with a robust maximum likelihood estimator (MLR) and full information maximum likelihood (FIML) to handle missing data (N = 7,903; people who answered at least one of the six items). Based on the item content, which distinguishes non-lethal applications (items 1--3: satellite image analysis, evacuation routing, communications disruption) from lethal applications (items 4--6: target identification, attack suggestions, autonomous destruction), we compared a one-factor model against a two-factor model separating non-lethal and lethal support.

Internal consistency pooled across all nine countries for the overall six-item scale was good ($\alpha$ = .90). The two subscales also showed satisfactory reliability (non-lethal: $\alpha$ = .85; lethal: $\alpha$ = .82). Country-specific reliability estimates are reported in Table \ref{tab:alpha_country}.

Both models showed acceptable fit. The one-factor model yielded CFI = .982, TLI = .970, RMSEA = .083, SRMR = .021, $\chi^2$(9, N = 7,903) = 226.73, p < .001. The two-factor model provided a marginally better fit: CFI = .987, TLI = .975, RMSEA = .076, SRMR = .018, $\chi^2$(8, N = 7,903) = 171.45, p < .001. A Satorra--Bentler scaled difference test confirmed that the two-factor model fit significantly better ($\Delta\chi^2$ = 52.18, $\Delta$df = 1, p < .001). Although the chi-square tests are significant for both models, this is expected given the large sample size. Table \ref{tab:cfa_loadings} presents the standardized factor loadings for both solutions.

\begin{table}[H]
\centering
\resizebox{\textwidth}{!}{
\begin{tabular}{p{0.45\textwidth} c c c}
\toprule
\textbf{Item Wording} & \textbf{1-Factor ($\lambda$)} & \textbf{Non-lethal ($\lambda$)} & \textbf{Lethal ($\lambda$)} \\
\midrule
The military uses AI to analyse satellite images and identify enemy supply routes. & 0.816 & 0.828 & \\[0.5em]
AI systems analyse current combat information and recommend safe evacuation routes. & 0.769 & 0.778 & \\[0.5em]
An AI system controls drones that specifically disrupt enemy communications. & 0.806 & 0.811 & \\[0.5em]
AI is used to identify potential targets for future air strikes. & 0.817 & & 0.828 \\[0.5em]
AI systems make suggestions for attack options based on real-time data during ongoing combat. & 0.816 & & 0.828 \\[0.5em]
An AI-controlled defence system automatically detects and destroys enemy units on the battlefield without human intervention. & 0.662 & & 0.673 \\[0.5em]
\bottomrule
\end{tabular}
}
\caption{Standardized factor loadings from confirmatory factor analyses of the military AI support scale. The one-factor model treats all six items as a single latent construct; the two-factor model distinguishes non-lethal (items 1--3) from lethal (items 4--6) applications.}
\label{tab:cfa_loadings}
\end{table}

However, the two subscales were very highly correlated at the observed level (mean non-lethal support and mean lethal support: $r = .77$, 95\% CI [.76, .78], $p < .001$), and the latent factor correlation was .96, indicating that the two dimensions are closely related and empirically difficult to separate. Given this high correlation and the finding that metric measurement invariance holds across all nine countries (see Section \ref{section:invariance}), we use the mean of all six items as a single support-for-military-AI index for our main models, as preregistered.

\subsection{Measurement invariance of Military AI use}
\label{section:invariance}

To assess whether the military AI support scale functions equivalently across the nine countries, we conducted a series of multigroup confirmatory factor analyses with increasingly restrictive equality constraints, using MLR estimation and FIML for missing data. We tested both the one-factor model and the two-factor model (non-lethal vs.\ lethal) at four levels of invariance: configural (same factor structure), metric (equal factor loadings), scalar (equal loadings and intercepts), and strict (equal loadings, intercepts, and residual variances). We used the change in robust CFI ($\Delta$CFI) as the primary criterion, following \cite{chen_sensitivity_2007}, where a decrease larger than .010 indicates a meaningful decline in fit.

Table \ref{tab:invariance} presents the fit indices for each level of invariance for both factor solutions.

\begin{table}[H]
\centering
\resizebox{\textwidth}{!}{
\begin{tabular}{llcccccc}
\toprule
Model & Level & CFI & TLI & RMSEA & SRMR & $\Delta$CFI & $\Delta$RMSEA \\
\midrule
\multirow{4}{*}{One-factor}
& Configural & .982 & .970 & .083 & .022 & --- & --- \\
& Metric     & .980 & .978 & .071 & .037 & -.002 & -.012 \\
& Scalar     & .965 & .971 & .081 & .049 & -.015 & +.011 \\
& Strict     & .940 & .961 & .094 & .061 & -.025 & +.012 \\
\addlinespace[4pt]
\multirow{4}{*}{Two-factor}
& Configural & .987 & .976 & .073 & .020 & --- & --- \\
& Metric     & .986 & .982 & .064 & .032 & -.001 & -.010 \\
& Scalar     & .977 & .977 & .072 & .040 & -.009 & +.008 \\
& Strict     & .958 & .969 & .083 & .051 & -.018 & +.011 \\
\bottomrule
\end{tabular}
}
\caption{Fit indices for measurement invariance tests of the military AI support scale across nine countries.}
\label{tab:invariance}
\end{table}

For both factor solutions, metric invariance was supported ($\Delta$CFI = -.002 for the one-factor model; $\Delta$CFI = -.001 for the two-factor model), indicating that the factor loadings are equivalent across countries. This level of invariance is sufficient for comparing regression coefficients across groups, as it ensures that the latent construct is measured on the same scale in each country \cite{chen_sensitivity_2007}.

Scalar invariance was not achieved for the one-factor model ($\Delta$CFI = -.015). For the two-factor model, the decline was borderline but remained just within the conventional threshold ($\Delta$CFI = -.009), suggesting approximate scalar invariance. This means that direct comparisons of latent means across countries should be interpreted with some caution, as observed score differences may partly reflect measurement artifacts rather than true differences in the underlying construct. For the present study, this limitation is of minor concern, as our main analyses focus on within-country associations (regression coefficients) rather than cross-national mean comparisons. The descriptive country-level means should therefore be treated as approximate.

\subsection{Predicting missing data}
\label{section:missing}
\begin{figure}[H]
    \centering
    \includegraphics[width=\linewidth]{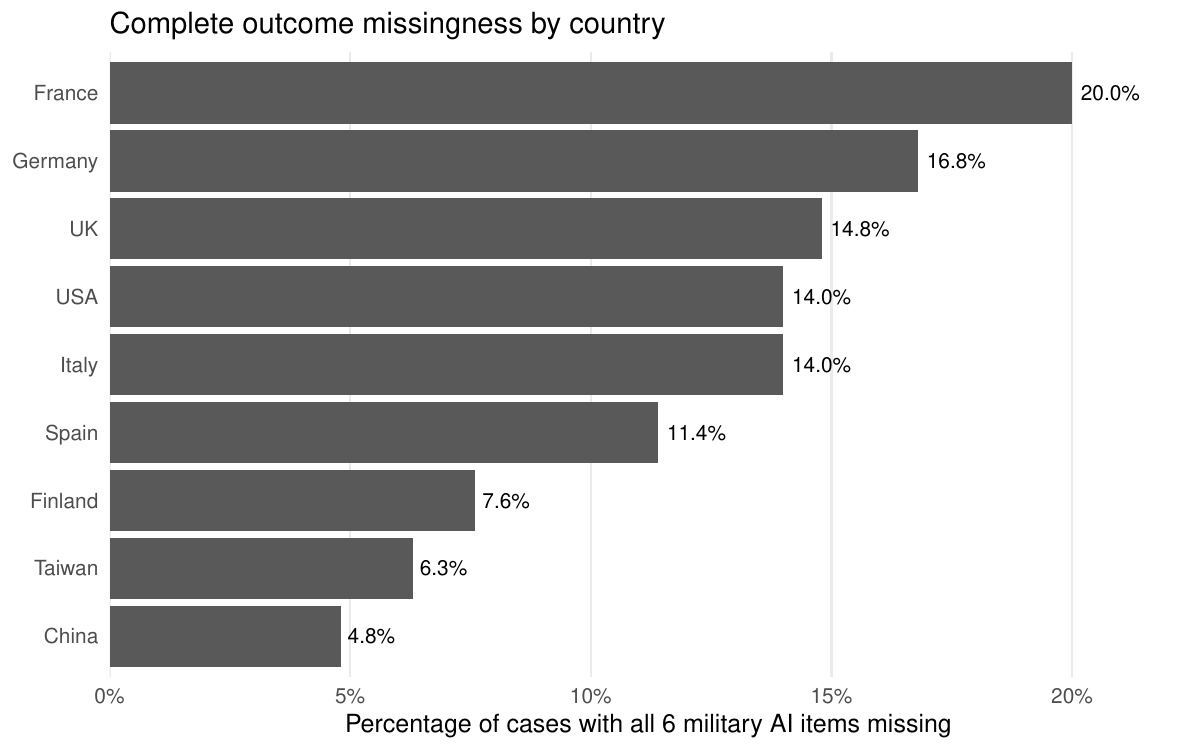}
    \caption{Complete outcome missingness by country. Bars represent the percentage of respondents with missing responses on all six military AI items. Countries are ordered by missingness rate.}
    \label{fig:missingness_country}
\end{figure}

\begin{figure}[H]
    \centering
    \includegraphics[width=\linewidth]{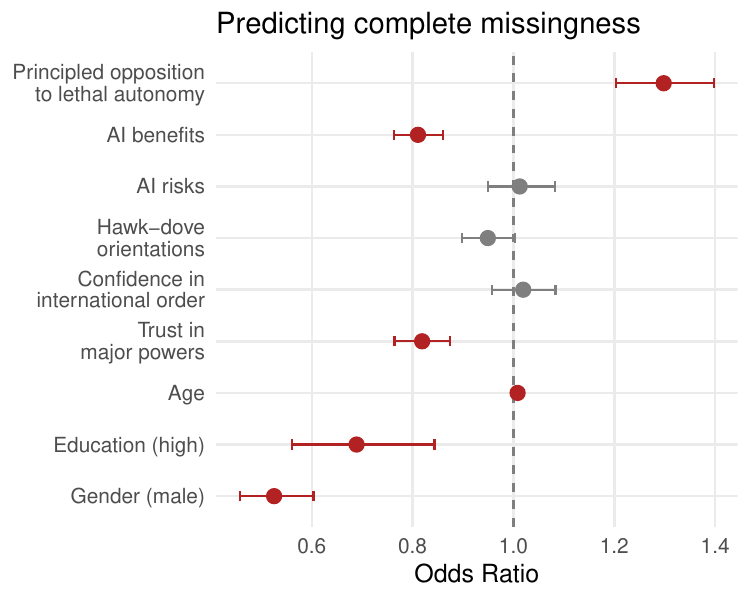}
    \caption{Forest plot of Odds Ratios from a Bayesian multilevel binary logistic regression predicting complete missingness for the six military AI use items. Points: posterior medians. Horizontal bars: 95 percent CIs. Dashed line: zero. Predictors whose CIs exclude zero are highlighted in red.}
    \label{fig:missingness_forest}
\end{figure}

As an additional analysis, we estimated a binary logistic Bayesian regression model with varying intercepts for countries (Figure~\ref{fig:missingness_forest}), testing which variables potentially explain complete missingness (none of the six items on military use of AI were answered). Rates of complete missingness varied across countries (Figure~\ref{fig:missingness_country}). Respondents could select "cannot assess" or simply not respond to an item. In all countries, complete missingness is primarily driven by "cannot assess" responses rather than item non-response. Outright non-response was low and similar across countries (1.3\%–2.9\%), whereas "cannot assess" rates varied widely, from 6.6\% in China and 7.2\% in Taiwan to 28.5\% in France. The non-responses we observed, therefore, do not appear to reflect unwillingness to express an opinion, but rather suggest that respondents have not yet formed strong views on the military use of AI.

 The analysis with complete missingness as the outcome shows several predictors of who fails to give a substantive answer to any of the six military AI items (Figure~\ref{fig:missingness_forest}). Older respondents, women, and those with lower education are each independently more likely to fall in this group. Stronger principled opposition to lethal autonomy is also associated with higher complete missingness, as are lower perceived benefits of AI in general and lower trust in major powers. Given that this missingness is driven predominantly by "cannot assess" rather than outright non-response, these results suggest that a specific demographic and attitudinal profile is associated with respondents who do not yet feel able to take a position on the military use of AI.

Because complete missingness is systematically related to respondents' demographics and attitudes, it cannot be treated as missing completely at random. Listwise deletion would therefore risk biasing estimates by under-representing exactly the profile of respondents identified above. To address this, our main analyses rely, as we preregistered, on multiple imputation rather than complete-case analysis; the imputation procedure is described in Section~\ref{model_results}. As a further check, we also re-estimated the main models on the subsample of respondents who provided at least one substantive answer on the military AI items (Section~\ref{sec:robustness_impute}); results are substantively unchanged.

\section{Model results}
\label{model_results}
In this section, we report all the models for the main results reported in the manuscript. For H1-H5 we used a Bayesian regression model with varying intercepts for countries. For RQ2 we estimated again a Bayesian regression model with varying intercepts for countries but this time with standardized predictors to make the effect strength comparable. For RQ3 we estimated as preregistered three different models: a baseline model with only varying intercepts, a model with varying intercepts and all predictors, and a model with varying intercepts and slopes for all predictors.

As specified in the preregistration under \textit{Missing data}, we used data imputation to fill in missing responses for all the predictors and the outcome variable. For data imputation, we followed the procedure recommended in the literature \cite{van-Buuren:2018aa}. Using the R package mice \cite{van-Buuren:2011aa}, we created 20 datasets with imputed data for the missing values using predictive mean matching \cite{van-Buuren:2018aa}. We used all measured variables on the item level as predictors for predictive mean matching. The mean indices were created after the data imputation. We then estimated the models for each dataset and pool the results with the brms package in R.

As a robustness analysis, we conducted an additional analysis, only with participants who at least provided an answer to one of the six items used for the outcome variable.  Imputation was applied to this restricted sample in the same way as in the main analysis. The results are substantively the same. The results of this analysis are reported in Section~\ref{sec:robustness_impute}. 

For the Bayesian regression models, we used the default priors in brms, as preregistered. All models were estimated using four chains, with 2,000 iterations per chain, including 1,000 warm-up iterations. For the model with varying slopes, we increased this to 4,000 iterations per chain, including 2,000 warm-up iterations. All models were estimated separately for each of the 20 imputed datasets, and the results were then pooled. The chains converged well across all models. For example, in the main model with varying intercepts, 422 of the 480 Rhat values were 1.00, and the remaining 58 were 1.01. Furthermore, multicollinearity is not a concern in the main model, as all variance inflation factors are low and range from 1.02 to 1.37, with VIFs of 1.37 for perceived AI benefits and 1.18 for perceived AI risks.

\subsection{Main hypotheses H1-H5}
\begin{table}[H]
\centering
\begin{tabular}[t]{lcc}
\toprule
Predictors & Estimate & 95\% CI \\
\midrule
Intercept & 0.84 & [0.61, 1.07] \\
H1: AI benefits index & 0.45 & [0.43, 0.47] \\
H2: AI risks index & 0.10 & [0.08, 0.12] \\
H3: Principled opposition index & $-$0.01 & [$-$0.03, 0.01] \\
H4: Hawk--dove index & 0.17 & [0.15, 0.19] \\
H5a: Confidence in intl.\ order index & 0.10 & [0.08, 0.12] \\
H5b: Trust in major powers index & $-$0.02 & [$-$0.04, $-$0.00] \\
Age & 0.01 & [0.01, 0.01] \\
Education (high) & $-$0.00 & [$-$0.06, 0.06] \\
Gender (male) & 0.10 & [0.06, 0.15] \\
\addlinespace
\multicolumn{3}{l}{\textit{Random Effects}} \\
$\sigma^2$ & 0.98 & \\
$\tau_{00}$ \textsubscript{country} & 0.04 & \\
ICC & 0.04 & \\
$N$ \textsubscript{country} & 9 & \\
Observations & 9000 & \\
Marginal $R^2$ / Conditional $R^2$ & 0.392 / 0.401 & \\
\bottomrule
\end{tabular}
\caption{Bayesian multilevel model predicting support for military AI with varying intercepts by country. Estimates
are posterior means with 95\% credible intervals from \texttt{brms}.}
\label{tab:h1h5}
\end{table}

\subsection{RQ2: Principled opposition vs general technology dispositions}
\label{sec:hypotheses_testing}

\begin{table}[H]
\centering
\begin{tabular}[t]{lcc}
\toprule
Predictors & Estimate & 95\% CI \\
\midrule
Intercept & 4.91 & [4.77, 5.06] \\
AI benefits index & 0.63 & [0.60, 0.66] \\
AI risks index & 0.13 & [0.10, 0.15] \\
Principled opposition index & $-$0.02 & [$-$0.04, 0.01] \\
Hawk--dove index & 0.23 & [0.21, 0.26] \\
Confidence in intl.\ order index & 0.13 & [0.10, 0.15] \\
Trust in major powers index & $-$0.03 & [$-$0.05, $-$0.00] \\
Age & 0.13 & [0.11, 0.16] \\
Education (high) & $-$0.00 & [$-$0.06, 0.06] \\
Gender (male) & 0.10 & [0.06, 0.15] \\
\addlinespace
\multicolumn{3}{l}{\textit{Random Effects}} \\
$\sigma^2$ & 0.98 & \\
$\tau_{00}$ \textsubscript{country} & 0.04 & \\
ICC & 0.04 & \\
$N$ \textsubscript{country} & 9 & \\
Observations & 9000 & \\
Marginal $R^2$ / Conditional $R^2$ & 0.382 / 0.391 & \\
\bottomrule
\end{tabular}
\caption{Bayesian multilevel model predicting support for military AI with varying intercepts by country. All predictors were standardized before estimating the model. Estimates are posterior means with 95\% credible intervals from \texttt{brms}.}
\label{tab:h1h5_rescaled}
\end{table}

\subsection{RQ3: Compositional vs structural differences}
We compare three models. The first model is a baseline model with only varying intercepts and no predictors. The second model includes all predictors and varying intercepts. The third model is with varying intercepts and slopes.

\begin{table}[H]
\centering
\begin{tabular}[t]{lcc}
\toprule
Predictors & Estimate & 95\% CI \\
\midrule
Intercept & 4.97 & [4.76, 5.17] \\
\addlinespace
\multicolumn{3}{l}{\textit{Random Effects}} \\
$\sigma^2$ & 1.57 & \\
$\tau_{00}$ \textsubscript{country} & 0.08 & \\
ICC & 0.05 & \\
$N$ \textsubscript{country} & 9 & \\
Observations & 9000 & \\
Marginal $R^2$ / Conditional $R^2$ & 0.000 / 0.030 & \\
\bottomrule
\end{tabular}
\caption{Bayesian null multilevel model predicting support for military AI with varying intercepts by country. Estimates are posterior means with 95\% credible intervals from \texttt{brms}.}
\label{tab:null_model}
\end{table}

\begin{table}[H]
\centering
\begin{tabular}[t]{lcc}
\toprule
Predictors & Estimate & 95\% CI \\
\midrule
Intercept & 0.80 & [0.59, 1.01] \\
H1: AI benefits index & 0.44 & [0.40, 0.48] \\
H2: AI risks index & 0.11 & [0.06, 0.15] \\
H3: Principled opposition index & $-$0.01 & [$-$0.06, 0.04] \\
H4: Hawk--dove index & 0.18 & [0.13, 0.22] \\
H5a: Confidence in intl.\ order index & 0.10 & [0.06, 0.14] \\
H5b: Trust in major powers index & $-$0.01 & [$-$0.05, 0.03] \\
Age & 0.01 & [0.01, 0.01] \\
Education (high) & $-$0.01 & [$-$0.07, 0.05] \\
Gender (male) & 0.10 & [0.05, 0.14] \\
\addlinespace
\multicolumn{3}{l}{\textit{Random Effects}} \\
$\sigma^2$ & 0.96 & \\
$\tau_{00}$ \textsubscript{country} & 0.02 & \\
$\tau_{11}$ \textsubscript{country.ben\_index} & 0.00 & \\
$\tau_{11}$ \textsubscript{country.ri\_index} & 0.00 & \\
$\tau_{11}$ \textsubscript{country.aut\_index} & 0.01 & \\
$\tau_{11}$ \textsubscript{country.dh\_index} & 0.00 & \\
$\tau_{11}$ \textsubscript{country.int\_index} & 0.00 & \\
$\tau_{11}$ \textsubscript{country.t\_index} & 0.00 & \\
ICC & 0.02 & \\
$N$ \textsubscript{country} & 9 & \\
Observations & 9000 & \\
Marginal $R^2$ / Conditional $R^2$ & 0.389 / 0.413 & \\
\bottomrule
\end{tabular}
\caption{Bayesian multilevel model predicting support for military AI use with varying intercepts and varying slopes by country. Estimates are posterior means with 95\% credible intervals from \texttt{brms}.}
\label{tab:varying_slopes_model}
\end{table}

\begin{table}[H]
\centering
\begin{tabular}{lrrrrrr}
\toprule
Model & elpd\_loo & SE & p\_loo & LOOIC & $\Delta$ELPD & SE$_{\Delta}$ \\
\midrule
Varying slopes and intercepts          & -12542.9 & 89.1 & 79.3 & 25085.8 & 0.0     & 0.0 \\
Varying intercepts with all predictors & -12591.1 & 88.0 & 26.3 & 25182.2 & -48.2   & 14.8 \\
Varying intercepts only                & -14783.2 & 82.3 & 11.8 & 29566.4 & -2240.3 & 71.9 \\
\bottomrule
\end{tabular}
\caption{LOO model comparison. Differences in expected log predictive density (ELPD) are reported relative to the best-performing model with varying slopes and intercepts. All Pareto $k$ estimates were below 0.7.}
\end{table}

\subsection{Robustness analysis with reduced sample}
\label{sec:robustness_impute}   
In this section, we report a model using a sample that includes only participants (n=7903) who have provided an answer to at least one of the six items used for the outcome variable. We use the same data imputation method as for the main analysis. The robustness model yields results that are highly similar to those of the main specification. The substantive conclusions are therefore unchanged.
\begin{table}[H]
\centering
\begin{tabular}[t]{lcc}
\toprule
Predictors & Estimate & 95\% CI \\
\midrule
Intercept & 0.68 & [0.45, 0.90] \\
H1: AI benefits index & 0.47 & [0.45, 0.49] \\
H2: AI risks index & 0.10 & [0.08, 0.12] \\
H3: Principled opposition index & $-$0.02 & [$-$0.04, 0.00] \\
H4: Hawk--dove index & 0.18 & [0.16, 0.20] \\
H5a: Confidence in intl.\ order index & 0.10 & [0.08, 0.12] \\
H5b: Trust in major powers index & $-$0.02 & [$-$0.03, 0.00] \\
Age & 0.01 & [0.01, 0.01] \\
Education (high) & 0.01 & [$-$0.06, 0.07] \\
Gender (male) & 0.12 & [0.07, 0.16] \\
\addlinespace
\multicolumn{3}{l}{\textit{Random Effects}} \\
$\sigma^2$ & 0.97 & \\
$\tau_{00}$ \textsubscript{country} & 0.04 & \\
ICC & 0.04 & \\
$N$ \textsubscript{country} & 9 & \\
Observations & 7903 & \\
Marginal $R^2$ / Conditional $R^2$ & 0.397 / 0.408 & \\
\bottomrule
\end{tabular}
\caption{Bayesian multilevel model predicting support for military AI with varying intercepts by country. Estimates are posterior means with 95\% credible intervals from \texttt{brms}.}
\label{tab:h1h5_full}
\end{table}

\subsection{Analysis of trust in major powers}
\label{major_powers}
We conducted one additional non-preregistered analysis using the trust in major powers items as single predictors rather than a mean index. In this analysis, instead of using a mean index, we use trust in the US, China, and Russia as single variables. We estimated the model with varying intercepts for countries and varying slopes for the disaggregated trust items.

The main model results are shown in Table~\ref{tab:h1h5_varying_trust}. The main interest of this additional analysis was the country-specific random slopes, which are shown in Table~\ref{tab:random_slopes_trust}. They are also used for Figure 5 in the main paper.
\begin{table}[H]
\centering
\begin{tabular}[t]{lcc}
\toprule
Predictors & Estimate & 95\% CI \\
\midrule
Intercept & 0.82 & [0.52, 1.12] \\
H1: AI benefits index & 0.45 & [0.42, 0.47] \\
H2: AI risks index & 0.11 & [0.09, 0.13] \\
H3: Principled opposition index & $-$0.01 & [$-$0.03, 0.01] \\
H4: Hawk--dove index & 0.17 & [0.15, 0.19] \\
H5a: Confidence in intl.\ order index & 0.10 & [0.08, 0.12] \\
H5b: Trust in the United States & 0.02 & [$-$0.01, 0.04] \\
H5c: Trust in China & $-$0.01 & [$-$0.03, 0.02] \\
H5d: Trust in Russia & $-$0.02 & [$-$0.05, 0.02] \\
Age & 0.01 & [0.01, 0.01] \\
Education (high) & $-$0.01 & [$-$0.07, 0.05] \\
Gender (male) & 0.10 & [0.06, 0.14] \\
\addlinespace
\multicolumn{3}{l}{\textit{Random Effects}} \\
$\sigma^2$ & 0.98 & \\
$\tau_{00}$ \textsubscript{country} & 0.12 & \\
$\tau_{11}$ \textsubscript{country.trust\_usa} & 0.00 & \\
$\tau_{11}$ \textsubscript{country.trust\_china} & 0.00 & \\
$\tau_{11}$ \textsubscript{country.trust\_russia} & 0.00 & \\
ICC & 0.11 & \\
$N$ \textsubscript{country} & 9 & \\
Observations & 9000 & \\
Marginal $R^2$ / Conditional $R^2$ & 0.382 / 0.396 & \\
\bottomrule
\end{tabular}
\caption{Bayesian multilevel model predicting support for military AI with varying intercepts by country and varying slopes for trust in the United States, China, and Russia. Estimates are posterior means with 95\% credible intervals from \texttt{brms}.}
\label{tab:h1h5_varying_trust}
\end{table}

\begin{table}[H]
\centering
\begin{tabular}[t]{llcc}
\toprule
Country & Trust variable & Slope & 95\% CI \\
\midrule
Germany & Trust USA & $-$0.002 & [$-$0.036, 0.030] \\
France & Trust USA & $-$0.003 & [$-$0.041, 0.028] \\
UK & Trust USA & 0.040 & [0.006, 0.074] \\
Italy & Trust USA & $-$0.004 & [$-$0.035, 0.027] \\
Spain & Trust USA & $-$0.006 & [$-$0.040, 0.024] \\
Finland & Trust USA & 0.018 & [$-$0.016, 0.052] \\
USA & Trust USA & 0.058 & [0.023, 0.094] \\
Taiwan & Trust USA & 0.015 & [$-$0.015, 0.045] \\
China & Trust USA & 0.023 & [$-$0.005, 0.051] \\
\addlinespace
Germany & Trust China & $-$0.006 & [$-$0.038, 0.026] \\
France & Trust China & $-$0.004 & [$-$0.034, 0.033] \\
UK & Trust China & $-$0.010 & [$-$0.049, 0.024] \\
Italy & Trust China & $-$0.009 & [$-$0.040, 0.022] \\
Spain & Trust China & $-$0.008 & [$-$0.039, 0.021] \\
Finland & Trust China & 0.006 & [$-$0.025, 0.052] \\
USA & Trust China & $-$0.001 & [$-$0.034, 0.038] \\
Taiwan & Trust China & $-$0.018 & [$-$0.057, 0.010] \\
China & Trust China & $-$0.007 & [$-$0.041, 0.026] \\
\addlinespace
Germany & Trust Russia & $-$0.043 & [$-$0.084, $-$0.005] \\
France & Trust Russia & $-$0.021 & [$-$0.061, 0.020] \\
UK & Trust Russia & 0.031 & [$-$0.014, 0.083] \\
Italy & Trust Russia & $-$0.047 & [$-$0.086, $-$0.009] \\
Spain & Trust Russia & $-$0.046 & [$-$0.088, $-$0.007] \\
Finland & Trust Russia & $-$0.049 & [$-$0.102, $-$0.003] \\
USA & Trust Russia & 0.014 & [$-$0.028, 0.056] \\
Taiwan & Trust Russia & $-$0.021 & [$-$0.061, 0.023] \\
China & Trust Russia & 0.021 & [$-$0.018, 0.063] \\
\bottomrule
\end{tabular}
\caption{Country-specific random slopes for trust in the United States, China, and Russia. Estimates are posterior means with 95\% credible intervals from the Bayesian multilevel model.}
\label{tab:random_slopes_trust}
\end{table}

\subsection{Robustness check using the highest-lethality military AI item}
\label{lethality_item}
We also estimated an additional non-preregistered model using the single item with the highest described lethality and autonomy level ("An AI-controlled defense system automatically detects and destroys enemy units on the battlefield without human intervention") to assess the stability of our analysis results using the six-item index. Overall, the results are broadly consistent with the index model. However, the variable principled opposition to lethal autonomy changes substantially (Table~\ref{tab:h1h5_single}). While this variable was not a substantial predictor for the index, it is clearly negatively associated with support for the highest-lethality item. This suggests that principled opposition is especially relevant when respondents evaluate the use of fully autonomous lethal force. At the same time, perceived AI benefits remain the strongest predictor of support, also in this more specific and high-lethality use case.

We further re-ran the standardized hypothesis tests from Section~\ref{sec:hypotheses_testing} using the autonomous-lethal item as the dependent variable (Table~\ref{tab:lethality_item_stand}). AI-benefit perceptions remain the strongest predictor of support ($\beta = 0.74$, 95\% CI [0.69, 0.78]), while categorical-moral opposition to lethal autonomy is now substantially negative ($\beta = -0.17$, 95\% CI [$-0.21$, $-0.13$]). In absolute terms, AI-benefit perceptions are decisively stronger than categorical-moral opposition to lethal autonomy ($\Delta|\beta| = 0.57$, 90\% CI [0.52, 0.62], BF $>$ 1000). AI-risk perceptions and categorical-moral opposition are similar in magnitude, with only weak evidence that AI-risk perceptions are larger ($\Delta|\beta| = 0.02$, 90\% CI [$-$0.01, 0.06], BF = 5.09).

\begin{table}[H]
\centering
\begin{tabular}[t]{lcc}
\toprule
Predictors & Estimate & 95\% CI \\
\midrule
Intercept & 0.76 & [0.41, 1.11] \\
H1: AI benefits index & 0.53 & [0.50, 0.56] \\
H2: AI risks index & 0.15 & [0.12, 0.19] \\
H3: Principled opposition index & $-$0.13 & [$-$0.16, $-$0.10] \\
H4: Hawk--dove index & 0.09 & [0.06, 0.12] \\
H5a: Confidence in intl.\ order index & 0.04 & [0.01, 0.07] \\
H5b: Trust in major powers index & 0.07 & [0.04, 0.10] \\
Age & 0.01 & [0.01, 0.01] \\
Education (high) & $-$0.14 & [$-$0.23, $-$0.04] \\
Gender (male) & 0.06 & [$-$0.00, 0.13] \\
\addlinespace
\multicolumn{3}{l}{\textit{Random Effects}} \\
$\sigma^2$ & 2.28 & \\
$\tau_{00}$ \textsubscript{country} & 0.10 & \\
ICC & 0.04 & \\
$N$ \textsubscript{country} & 9 & \\
Observations & 9000 & \\
Marginal $R^2$ / Conditional $R^2$ & 0.253 / 0.276 & \\
\bottomrule
\end{tabular}
\caption{Bayesian multilevel model predicting support for the highest-lethality military AI use item with varying intercepts by country. Estimates are posterior means with 95\% credible intervals from \texttt{brms}.}
\label{tab:h1h5_single}
\end{table}

\begin{table}[H]
\centering
\begin{tabular}[t]{lcc}
\toprule
Predictors & Estimate & 95\% CI \\
\midrule
Intercept & 4.44 & [4.21, 4.67] \\
H1: AI benefits index & 0.74 & [0.69, 0.78] \\
H2: AI risks index & 0.19 & [0.15, 0.23] \\
H3: Principled opposition index & $-$0.17 & [$-$0.21, $-$0.13] \\
H4: Hawk--dove index & 0.13 & [0.09, 0.16] \\
H5a: Confidence in intl.\ order index & 0.05 & [0.01, 0.09] \\
H5b: Trust in major powers index & 0.11 & [0.07, 0.15] \\
Age & 0.15 & [0.12, 0.19] \\
Education (high) & $-$0.14 & [$-$0.23, $-$0.04] \\
Gender (male) & 0.06 & [$-$0.00, 0.13] \\
\addlinespace
\multicolumn{3}{l}{\textit{Random Effects}} \\
$\sigma^2$ & 2.28 & \\
$\tau_{00}$ \textsubscript{country} & 0.10 & \\
ICC & 0.04 & \\
$N$ \textsubscript{country} & 9 & \\
Observations & 9000 & \\
Marginal $R^2$ / Conditional $R^2$ & 0.253 / 0.276 & \\
\bottomrule
\end{tabular}
\caption{Robustness check predicting support for the highest-lethality military AI use item with varying intercepts by country. All predictors were standardized before estimating the model. Estimates are posterior means with 95\% credible intervals from \texttt{brms}.}
\label{tab:lethality_item_stand}
\end{table}

\subsection{Robustness check using two separate dimensions of military AI use}
In this section, we also report Bayesian regression models for each dimension of military use of AI. We estimate one model with an index comprising the three items describing non-lethal use (Table~\ref{tab:non_lethal}), and another model with an index comprising the three items describing lethal use of AI (Table~\ref{tab:lethal}).

Results are broadly similar across both sub-indices, with one substantively important exception. Categorical-moral opposition to lethal autonomy is positively associated with support for non-lethal applications (β = 0.03, 95\% CI [0.01, 0.06]) but negatively associated with support for lethal applications (β = -0.06, 95\% CI [-0.08, -0.04]). Together with the stronger negative association observed at the autonomous-lethal item (β = -0.17; Section~\ref{lethality_item}), this gradient shows that the autonomous-weapons frame is concentrated in applications involving lethal force and autonomy rather than absent in the application space. AI-benefit perceptions and hawk-dove orientations remain the largest predictors in both sub-indices, with the hawk-dove somewhat smaller for the lethal sub-index.

\begin{table}[H]
\centering
\begin{tabular}[t]{lcc}
\toprule
Predictors & Estimate & 95\% CI \\
\midrule
Intercept & 0.73 & [0.46, 1.00] \\
H1: AI benefits index & 0.49 & [0.46, 0.51] \\
H2: AI risks index & 0.11 & [0.09, 0.14] \\
H3: Principled opposition index & $-$0.06 & [$-$0.08, $-$0.04] \\
H4: Hawk--dove index & 0.17 & [0.14, 0.19] \\
H5a: Confidence in intl.\ order index & 0.08 & [0.06, 0.10] \\
H5b: Trust in major powers index & 0.01 & [$-$0.01, 0.03] \\
Age & 0.01 & [0.01, 0.01] \\
Education (high) & $-$0.05 & [$-$0.12, 0.02] \\
Gender (male) & 0.09 & [0.04, 0.14] \\
\addlinespace
\multicolumn{3}{l}{\textit{Random Effects}} \\
$\sigma^2$ & 1.21 & \\
$\tau_{00}$ \textsubscript{country} & 0.07 & \\
ICC & 0.05 & \\
$N$ \textsubscript{country} & 9 & \\
Observations & 9000 & \\
Marginal $R^2$ / Conditional $R^2$ & 0.369 / 0.385 & \\
\bottomrule
\end{tabular}
\caption{Bayesian multilevel model predicting support for lethal military AI use with varying intercepts by country. Estimates are posterior means with 95\% credible intervals from \texttt{brms}.}
\label{tab:lethal}
\end{table}

\begin{table}[H]
\centering
\begin{tabular}[t]{lcc}
\toprule
Predictors & Estimate & 95\% CI \\
\midrule
Intercept & 0.95 & [0.73, 1.17] \\
H1: AI benefits index & 0.41 & [0.39, 0.44] \\
H2: AI risks index & 0.10 & [0.07, 0.12] \\
H3: Principled opposition index & 0.03 & [0.01, 0.06] \\
H4: Hawk--dove index & 0.18 & [0.16, 0.20] \\
H5a: Confidence in intl.\ order index & 0.11 & [0.09, 0.14] \\
H5b: Trust in major powers index & $-$0.05 & [$-$0.07, $-$0.03] \\
Age & 0.01 & [0.01, 0.01] \\
Education (high) & 0.04 & [$-$0.03, 0.11] \\
Gender (male) & 0.11 & [0.07, 0.16] \\
\addlinespace
\multicolumn{3}{l}{\textit{Random Effects}} \\
$\sigma^2$ & 1.12 & \\
$\tau_{00}$ \textsubscript{country} & 0.02 & \\
ICC & 0.02 & \\
$N$ \textsubscript{country} & 9 & \\
Observations & 9000 & \\
Marginal $R^2$ / Conditional $R^2$ & 0.331 / 0.333 & \\
\bottomrule
\end{tabular}
\caption{Bayesian multilevel model predicting support for non-lethal military AI use with varying intercepts by country. Estimates are posterior means with 95\% credible intervals from \texttt{brms}.}
\label{tab:non_lethal}
\end{table}

\begin{table}[H]
\centering
\begin{tabular}[t]{lcc}
\toprule
Predictors & Estimate & 95\% CI \\
\midrule
Intercept & 4.75 & [4.56, 4.93] \\
H1: AI benefits index & 0.68 & [0.65, 0.71] \\
H2: AI risks index & 0.14 & [0.11, 0.17] \\
H3: Principled opposition index & $-$0.08 & [$-$0.10, $-$0.05] \\
H4: Hawk--dove index & 0.22 & [0.19, 0.25] \\
H5a: Confidence in intl.\ order index & 0.11 & [0.08, 0.14] \\
H5b: Trust in major powers index & 0.02 & [$-$0.01, 0.05] \\
Age & 0.13 & [0.11, 0.16] \\
Education (high) & $-$0.05 & [$-$0.12, 0.02] \\
Gender (male) & 0.09 & [0.04, 0.14] \\
\addlinespace
\multicolumn{3}{l}{\textit{Random Effects}} \\
$\sigma^2$ & 1.21 & \\
$\tau_{00}$ \textsubscript{country} & 0.07 & \\
ICC & 0.05 & \\
$N$ \textsubscript{country} & 9 & \\
Observations & 9000 & \\
Marginal $R^2$ / Conditional $R^2$ & 0.368 / 0.384 & \\
\bottomrule
\end{tabular}
\caption{Bayesian multilevel model predicting support for lethal military AI use with varying intercepts by country. All predictors were standardized before estimating the model. Estimates are posterior means with 95\% credible intervals from \texttt{brms}.}
\label{tab:lethal_stand}
\end{table}

\begin{table}[H]
\centering
\begin{tabular}[t]{lcc}
\toprule
Predictors & Estimate & 95\% CI \\
\midrule
Intercept & 5.08 & [4.97, 5.20] \\
H1: AI benefits index & 0.58 & [0.55, 0.61] \\
H2: AI risks index & 0.12 & [0.09, 0.14] \\
H3: Principled opposition index & 0.04 & [0.01, 0.07] \\
H4: Hawk--dove index & 0.24 & [0.22, 0.27] \\
H5a: Confidence in intl.\ order index & 0.15 & [0.12, 0.18] \\
H5b: Trust in major powers index & $-$0.08 & [$-$0.11, $-$0.05] \\
Age & 0.13 & [0.11, 0.16] \\
Education (high) & 0.04 & [$-$0.02, 0.11] \\
Gender (male) & 0.11 & [0.07, 0.16] \\
\addlinespace
\multicolumn{3}{l}{\textit{Random Effects}} \\
$\sigma^2$ & 1.12 & \\
$\tau_{00}$ \textsubscript{country} & 0.02 & \\
ICC & 0.02 & \\
$N$ \textsubscript{country} & 9 & \\
Observations & 9000 & \\
Marginal $R^2$ / Conditional $R^2$ & 0.330 / 0.332 & \\
\bottomrule
\end{tabular}
\caption{Bayesian multilevel model predicting support for non-lethal military AI use with varying intercepts by country. All predictors were standardized before estimating the model. Estimates are posterior means with 95\% credible intervals from \texttt{brms}.}
\label{tab:non_lethal_stand}
\end{table}

\printbibliography[
  heading=bibintoc,
  title={Supplementary References}
]

\end{refsection}

\end{document}